\documentclass[namedreferences]{SolarPhysics}
\usepackage[optionalrh]{spr-sola-addons} 
\usepackage{graphicx}        
\usepackage{color}           
\usepackage{url}             

\newcommand{\beq}{\begin{equation}}
\newcommand{\eeq}{\end{equation}}
\newcommand{\beqa}{\begin{eqnarray}}
\newcommand{\eeqa}{\end{eqnarray}}

\renewcommand{\vec}[1]{ {\mathbf #1} }

\begin{document}

\begin{article}

\begin{opening}

\title{Multiwavelength Observations of Supersonic Plasma Blob Triggered by Reconnection Generated Velocity Pulse in AR10808}

\author{A.K.~\surname{Srivastava}$^{1,2}$\sep
    R.~\surname{Erd\'elyi}$^{2}$\sep
    K.~\surname{Murawski}$^{3}$\sep
    Pankaj~\surname{Kumar}$^{4}$\sep
}
    
\runningauthor{Srivastava et al.}
\runningtitle{Supersonic Plasma Blob}
\institute{$^{1,2}$ Aryabhatta Research Institute of Observational Sciences (ARIES), Nainital, India.
                     email: \url{aks@aries.res.in}\\
             $^{2}$Solar Physics and Space Plasma Research Centre (SP$^{2}$RC), School of Mathematics and Statistics, The University of Sheffield, Sheffield, U.K. \\
$^{3}$Group of Astrophysics,
             UMCS, ul. Radziszewskiego 10, 20-031 Lublin, Poland. \\
$^{4}$Korea Astronomy and Space Science Institute (KASI), Daejeon, 305-348, Republic of Korea.}

\begin{abstract}
Using multi-wavelength observations of Solar and Heliospheric Observatory (SoHO)/Michelson Doppler Imager (MDI),
Transition Region and Coronal
Explorer (TRACE) 171 \AA, and H$\alpha$ from Culgoora Solar Observatory at Narrabri, Australia,
we present a unique observational signature of a propagating 
supersonic plasma blob before an M6.2 class solar flare 
in AR10808 on 9th September 2005. The blob was observed between
05:27 UT to 05:32 UT with almost a constant shape 
for the first 2-3 minutes, and thereafter it quickly vanished in the corona. The 
observed lower bound speed of the blob is estimated as $\sim$215 km s$^{-1}$
in its dynamical phase. 
The evidence of the blob with almost similar shape and velocity concurrent in H$\alpha$
and TRACE 171 \AA\ supports its formation by multi-temperature plasma. 
The energy release by a recurrent 3-D reconnection process via the separator dome below the magnetic 
null point, between the emerging flux and pre-existing field lines in the lower solar atmosphere,
is found to be the driver of  a radial velocity pulse outwards that
accelerates this 
plasma blob in the solar atmosphere. 
In support of identification of the possible driver of the observed eruption,
we solve the two-dimensional ideal magnetohydrodynamic equations numerically 
to simulate the observed supersonic plasma blob.
The numerical modelling closely match the observed velocity,
evolution of multi-temperature plasma, and quick vanishing of the blob found in the observations.
Under typical coronal conditions, such blobs may also carry an energy flux of 
7.0$\times$10$^{6}$
ergs cm$^{-2}$ s$^{-1}$ to re-balance the coronal losses above active regions.
\end{abstract}
\end{opening}

\section{Introduction}
Large-scale solar transient phenomena 
(solar flares and CMEs) are examples of the magnetic dynamical 
and highly energetic processes of the Sun during which the energy stored in the 
sheared and twisted magnetic structures is rapidly released in 
the form of particle acceleration, plasma heating and bulk 
mass motion within from a few minutes to a few tens of minutes timescales. However,
the exact
mechanism of the initiation and triggering of solar flares and coronal
mass ejections (CMEs) since the discovery of the first flare 
by R.C. Carrington on 1st September, 1859
is still not yet known in full details.
It is well explored that the flares may occur near the magnetic null points, and current 
sheets are formed by instability near the neutral point 
where field lines exhibit a magnetic reconnection process. Reconnection 
can be an efficient mechanism of converting magnetic energy to 
thermal and bulk kinetic energies and to accelerate particles \cite{benz2008}. 
The novel observational
signature of 3-D X-type loop-loop interaction and the flare triggering
due to reconnection in an active magnetic complex has recently been 
reported by \inlinecite{kumar2010a}.
It is also found that the instabilities (kink, magnetic Kelvin-Helmholtz, coalescence, etc.) can also be
one of the favourable mechanisms to trigger solar flares 
with or without CMEs \cite{liu2007,sri2010a,kumar2010b,Claire2011,Botha12}. In conclusion,
appropriately correlated theoretical and observational studies 
of these phenomena are still needed to be explored to understand the complex dynamical 
processes at the Sun. 

In spite of a number of attempts to analyse the energy build-up and energy release processes
associated with the eruptive phenomena, the recent trend has also
been evolved to study the responses of solar eruptions on the plasma
dynamics and the generation of periodic (quasi-periodic) oscillation patterns 
during these energetic events. These natural responses 
to the highly energetic solar eruptive phenomena are very useful in
diagnosing and constraining the local plasma conditions of
the solar atmosphere. The high-speed 
plasmoids associated
with CME eruptions \cite{sheel2002,mano2003,lin2005,riley2007,milligan2010}, 
chromospheric evaporation \cite{veronig2010}, very high-speed outflows 
\cite{wang2006}, down-flows and plasma condensations \cite{yokoyama2001} in the
course of flaring activities have been extensively studied.
The secondary plasmoids can also be evolved in the 
eruptive region due to the plasmoid instability in a 
sufficiently long and thin current layer \cite{Lu07}.
Therefore, they are associated with
reconnection, however, not directly with CMEs. The ejecta can only generate
the favorable conditions for plasmoid instability by formation of elongated and
stretched current sheet in the active region.
Periodic and quasi-periodic MHD oscillations have also been observed and studied
as a natural response of the solar flares in their vicinity regions 
\cite{roberts1983,asc1999,wang2002,foullon2005,Istvan2005,nakariakov2006,Youra2007,Erd08}.
The energy deposited in these MHD oscillations are found to 
be orders of magnitude smaller compared to the flare energy release \cite{terradas2007}.
On the other hand, flare and CME induced waves and oscillations are also very useful to gain first insight into the 
local plasma conditions and dynamical processes of the active regions
\cite{Erd07,andries2009,asc2009,Tar09,terradas2009,ofman2009,ruderman2009}. 

In addition to the large-scale motions in terms of flows and ejecta, waves and oscillations of 
magneto-fluids, and energetics of solar eruptions, the formation of small-scale 
plasma motions, e.g., anemone jets, spicules, penumbral jets 
\cite{depontieu2004,Bart06,depontieu2007,katsukawa2007,Ku11}, 
waves in form of transient pulses and 
wave-packets \cite{zaqarashvili2010,murawski2010,sri2010b} are important 
in observations to understand the heating and dynamics of the solar plasma at small 
spatio-temporal scales. These MHD/HD pulses may also be important 
to power small-scale transients in the lower solar atmosphere \cite{murawski2010}.
The theory of MHD pulses and wave excitation are somewhat established 
\cite{roberts1979,carlsson1997,selwa2010,selwa2007,murawski2010}. However, 
observational signatures are not very abundant in past few years 
to understand their role in the excitation of solar transients (e.g., jets
at various spatio-temporal scales). On the other hand, the role
of magnetic reconnection, twisted flux emergence from sub-photospheric layers,
and generation of instabilities are well-studied both in observations 
and theories in the jet productive regions \cite{Len07,Nis08,Bor09,Par09}.  

In the present paper, we analyze multi-wavelength 
imaging data of the Active Region NOAA AR 10808
that has produced violent explosions in the form of giant solar flares and associated eruptions. 
However, this active region has exhibited a very interesting phenomena, 
surprisingly rather quietly before an M-class flare on 9 September 2005, in the form of a supersonic
cylindrical plasma blob ejection from the flare site between 05:27 UT and 5:33 UT. 
The blob was observed by TRACE 171 \AA\ and also in H$_{\alpha}$ image sequences captured by Culgoora 
Solar Observatory, Australia. We present the observational data in Sec.~2. A numerical
modelling and the theoretical interpretations of the supersonic plasma blob is described in Sec.~3. The results and discussions are given in the last section.
\begin{figure}
\centering
\mbox{
\includegraphics[width=5.5cm, angle=90]{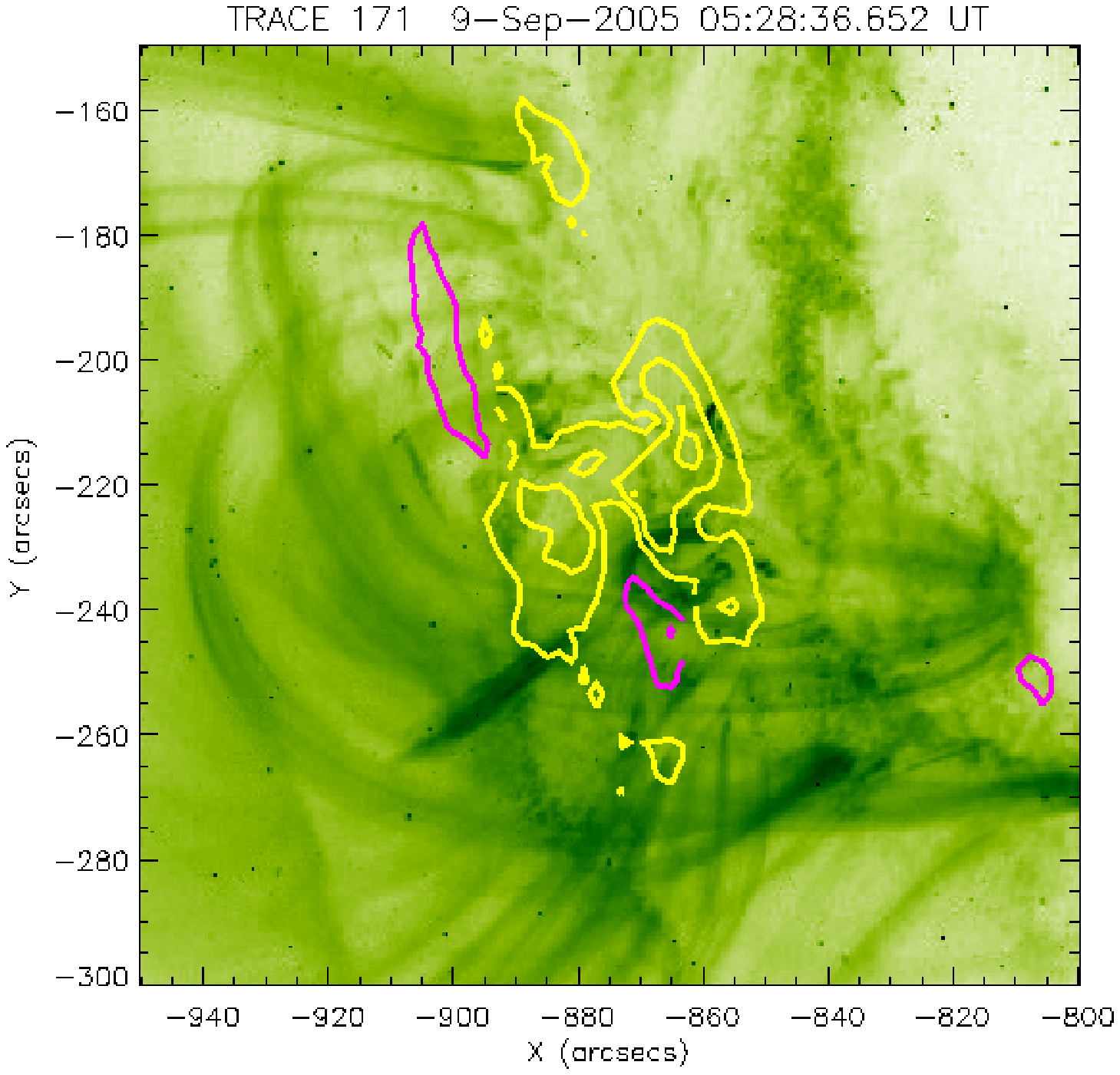}
\includegraphics[width=5.5cm, angle=90]{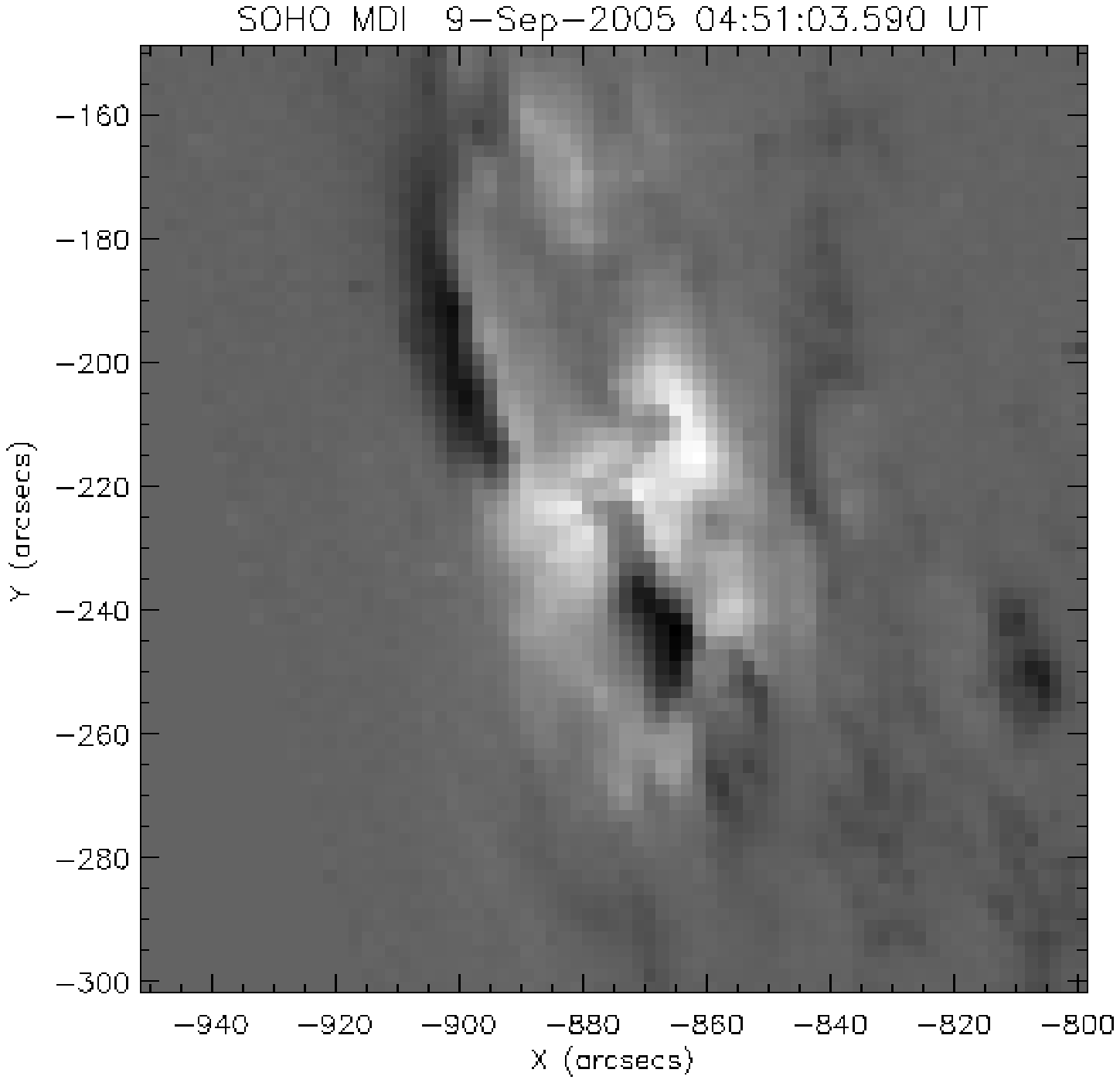}
}
\caption{Field-of-views : Co-aligned MDI contours overlaid on TRACE 171 \AA\
context image of the Active Region 10808 at 05:28:36 UT on 9 September, 2005 (left-panel).
Yellow (magenta) contours show the positive (negative) polarity of the photospheric magnetic field.
The maximum level of the magnetic field contours are $\pm$4000 G. For the positive polarity,
the two concentric iso-contours of magnetic fields are visible.
The plasma blob is shown to be propagating up in the solar atmosphere from its origin site 
associated with maximum magnetic fields.
The observed configuration of the active region magnetic field is shown in the right panel on 04:51 UT,
which later comes more inside the disk during the transient event.
}
\label{fig1}
\end{figure}
%
\section{Observations of Supersonic Plasma Blob}
\label{obs}
Active Region NOAA 10808 has appeared on the east 
limb on 07 September 2005 and produced at least 10 X-class 
and 25 M-class solar flares until it disappeared behind 
the solar-disk (Nagashima et al., 2007). The active 
region was located nearby the eastern limb of the southern 
hemisphere near the equatorial plane at S10E67 with $\beta\gamma\delta$ field 
configuration of the sunspot group. This active region was also
associated with CMEs and filament eruptions, 
and produced numerous space weather activities 
by its furious eruptive events
\cite{archontis2010,canou2009,nagashima2007,wang2006,li2007}. 
Figure 1 shows the coronal
image of AR 10808 as observed from the TRACE 171 \AA\ filter at 05:28:36 UT on 09 September, 2005.
The TRACE field of view is overlaid by co-aligned MDI
contours showing positive (yellow) and negative
(magenta) magnetic polarities (left panel of Figure 1). The positive and negative polarities
seem to be elongated respectively towards north-west to south-east, and north-east to south-west, with  tongue-like
structures (cf., right panel of Figure 1). This is a very typical signature of AR10808 and seems to be the evidence of
the emergence of twisted rising flux \cite{archontis2010,li2007}.
This typical characteristic of the AR10808 was one of the main causes that were 
responsible for the very energetic flares, CMEs, and filament eruptions, though
the rotation of spots and the tilt of the bipolar weightage axis were also
observed in this active region \cite{archontis2010,canou2009,nagashima2007,wang2006,li2007}. 
Although the violent solar eruptions have been extensively 
analyzed from this active region, we highlight here
an additional and an interesting plasma dynamics during an M-class flare that took place 
for a short duration of about 5-7 minutes 
on 9 September 2005. The M6.2 class solar flare has occurred in AR10808 at the location
S10E66. It starts at 05:32 UT, peaks at 05:48 UT, and ended at 06:00 UT.
The observed short duration plasma dynamics that is the interest and theme of this paper, has occurred well before this M6.2 class
solar flare just at the flare brightening site.
\begin{figure*}
\centering
\mbox{
\includegraphics[width=5cm]{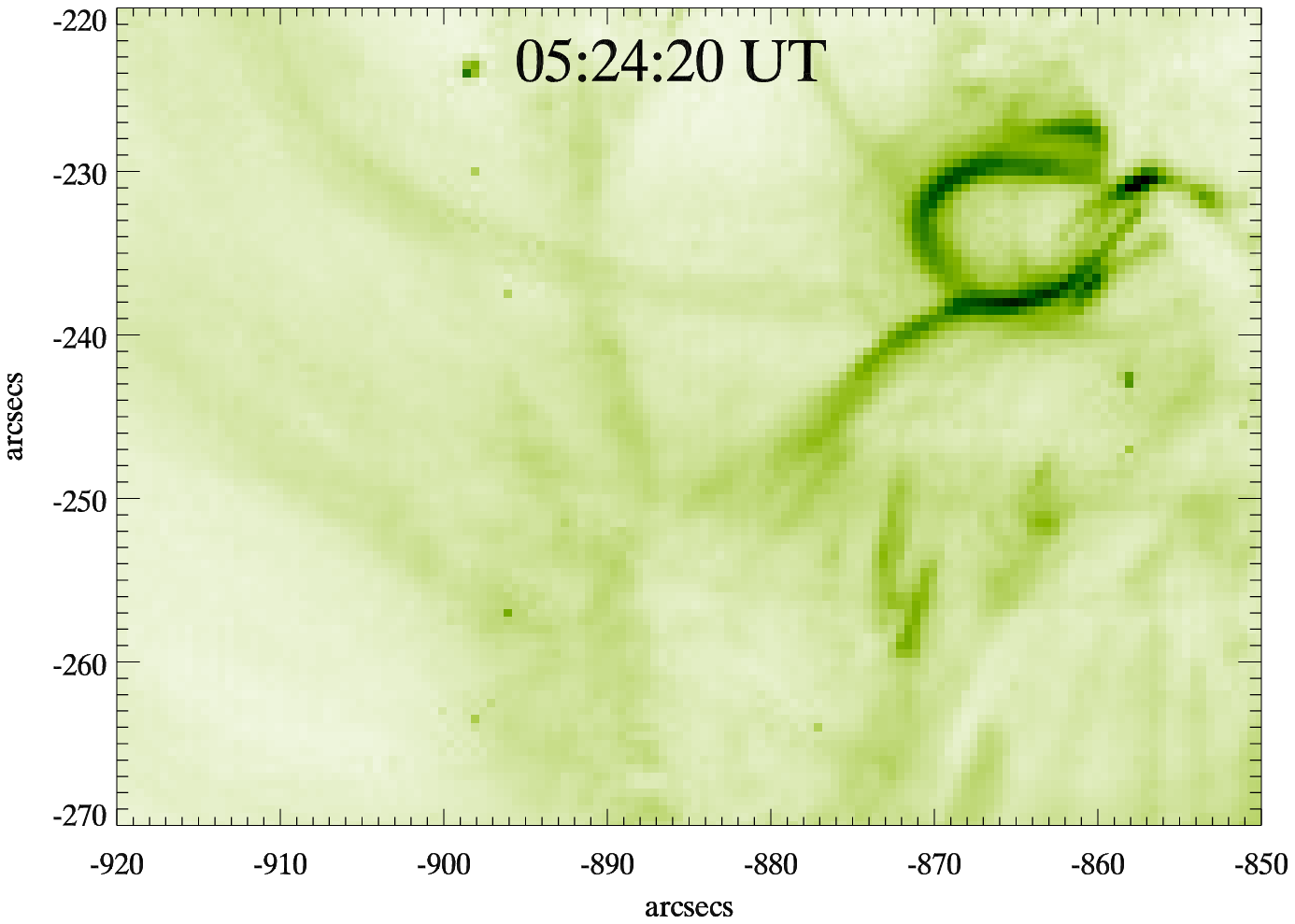}
\includegraphics[width=5cm]{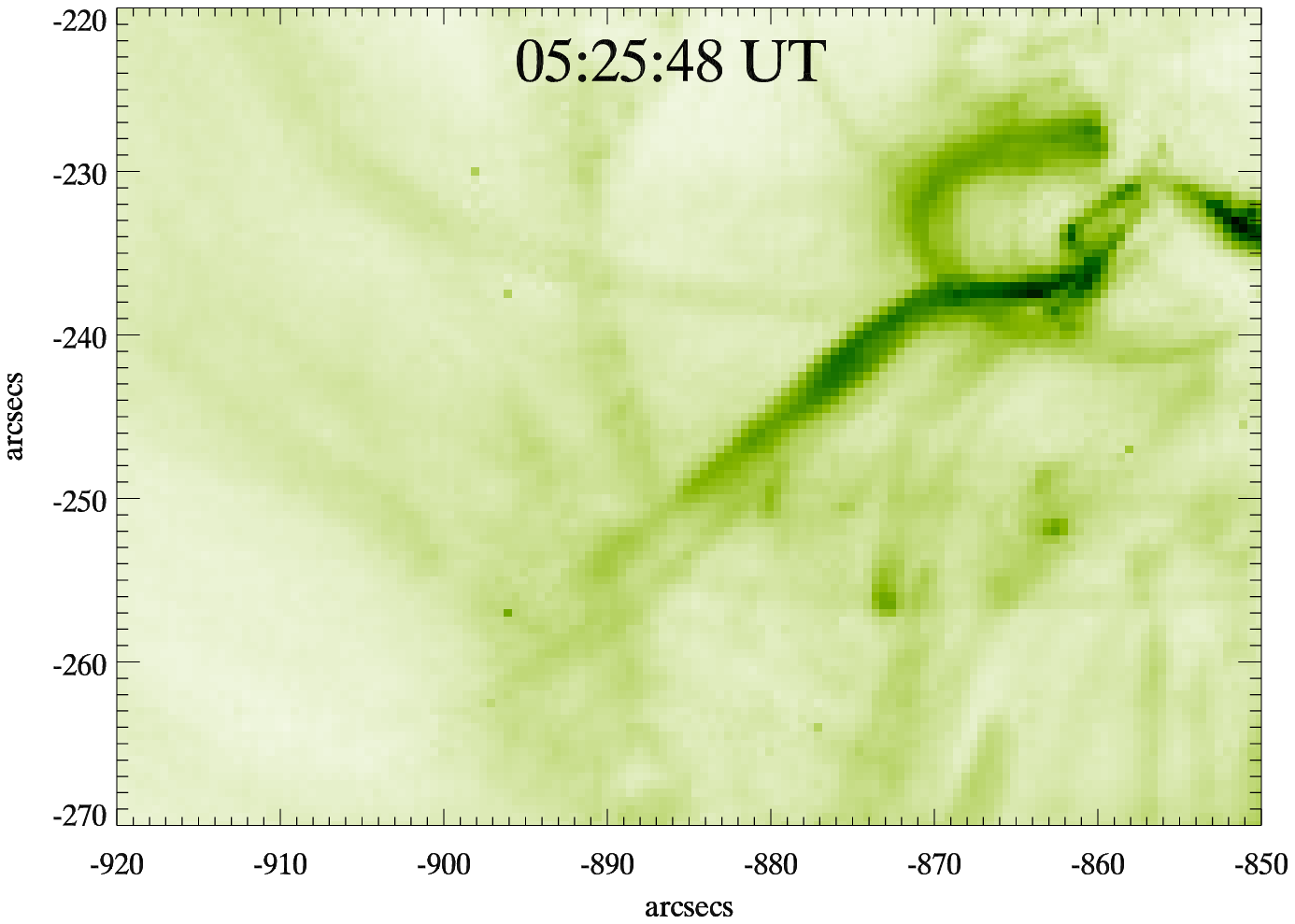}}
\mbox{
\includegraphics[width=5cm]{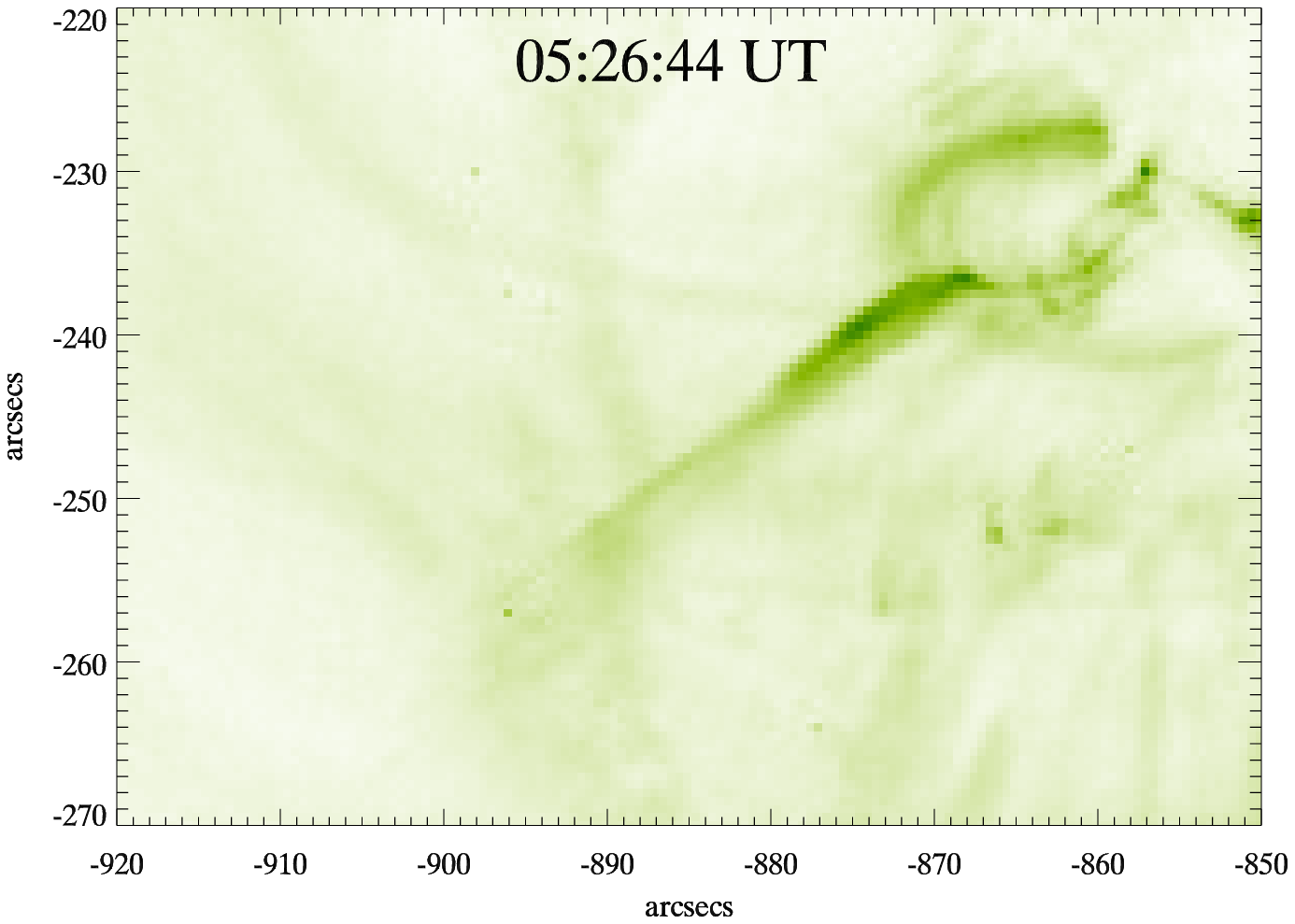}
\includegraphics[width=5cm]{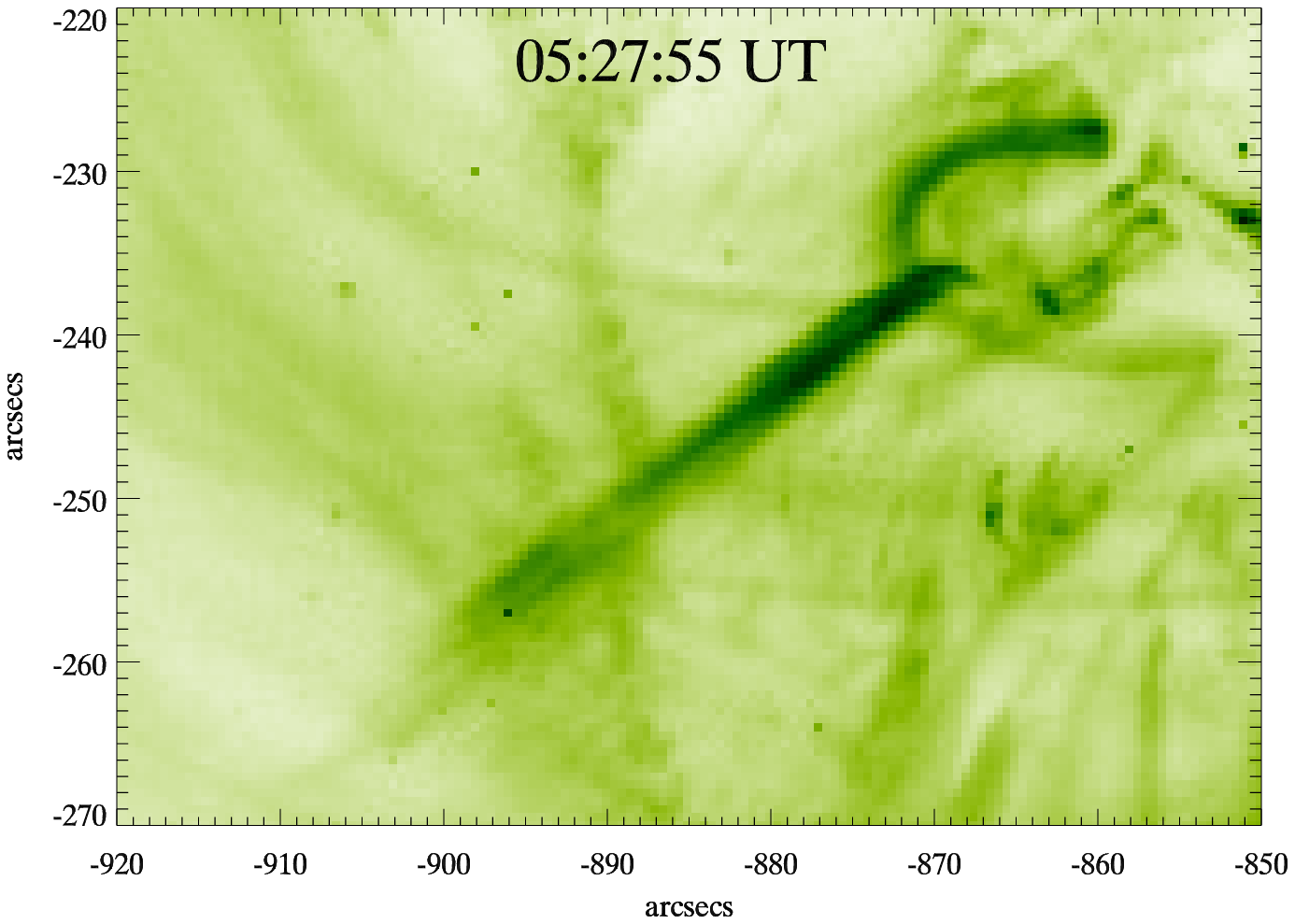}}
\mbox{
\includegraphics[width=5cm]{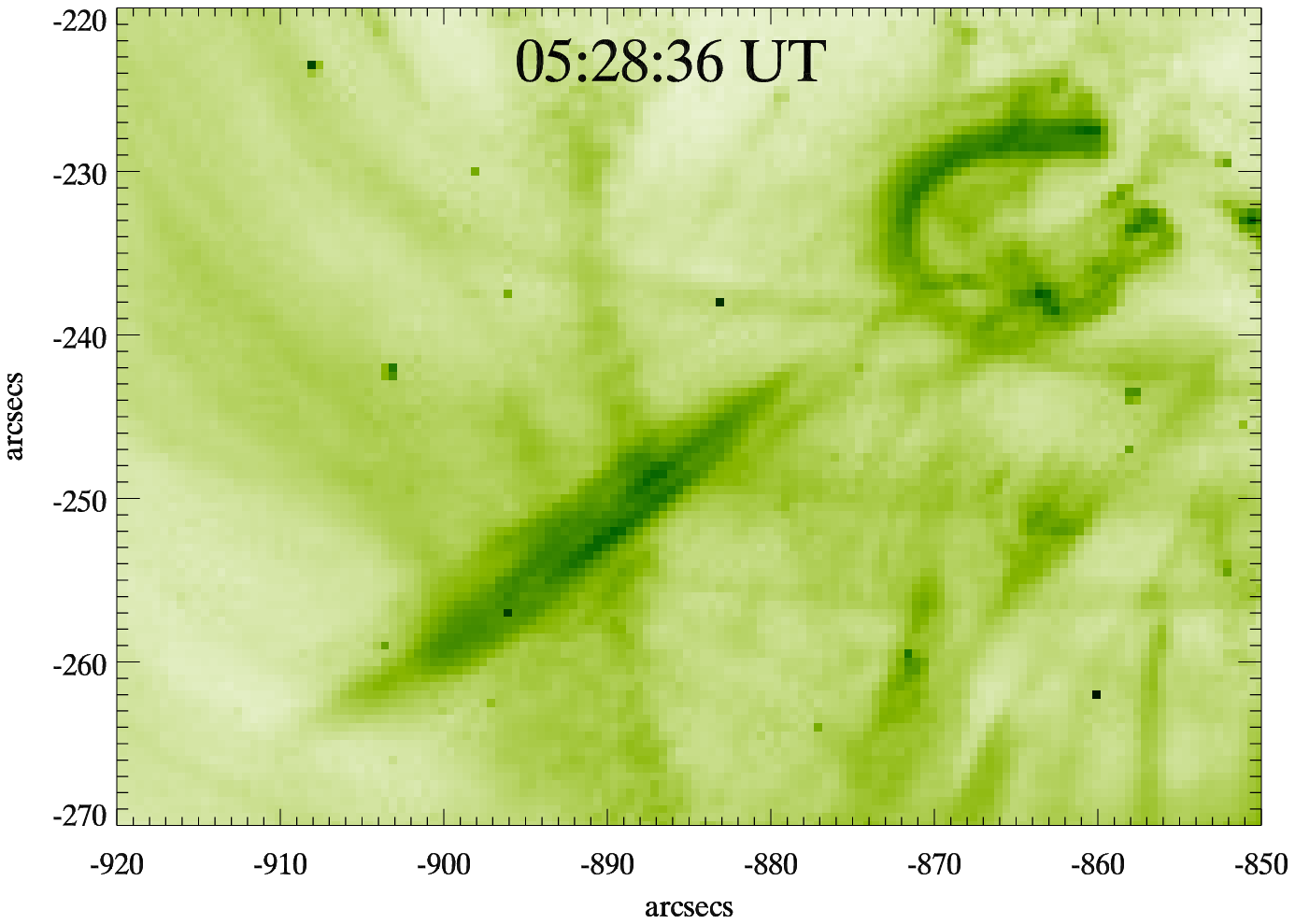}
\includegraphics[width=5cm]{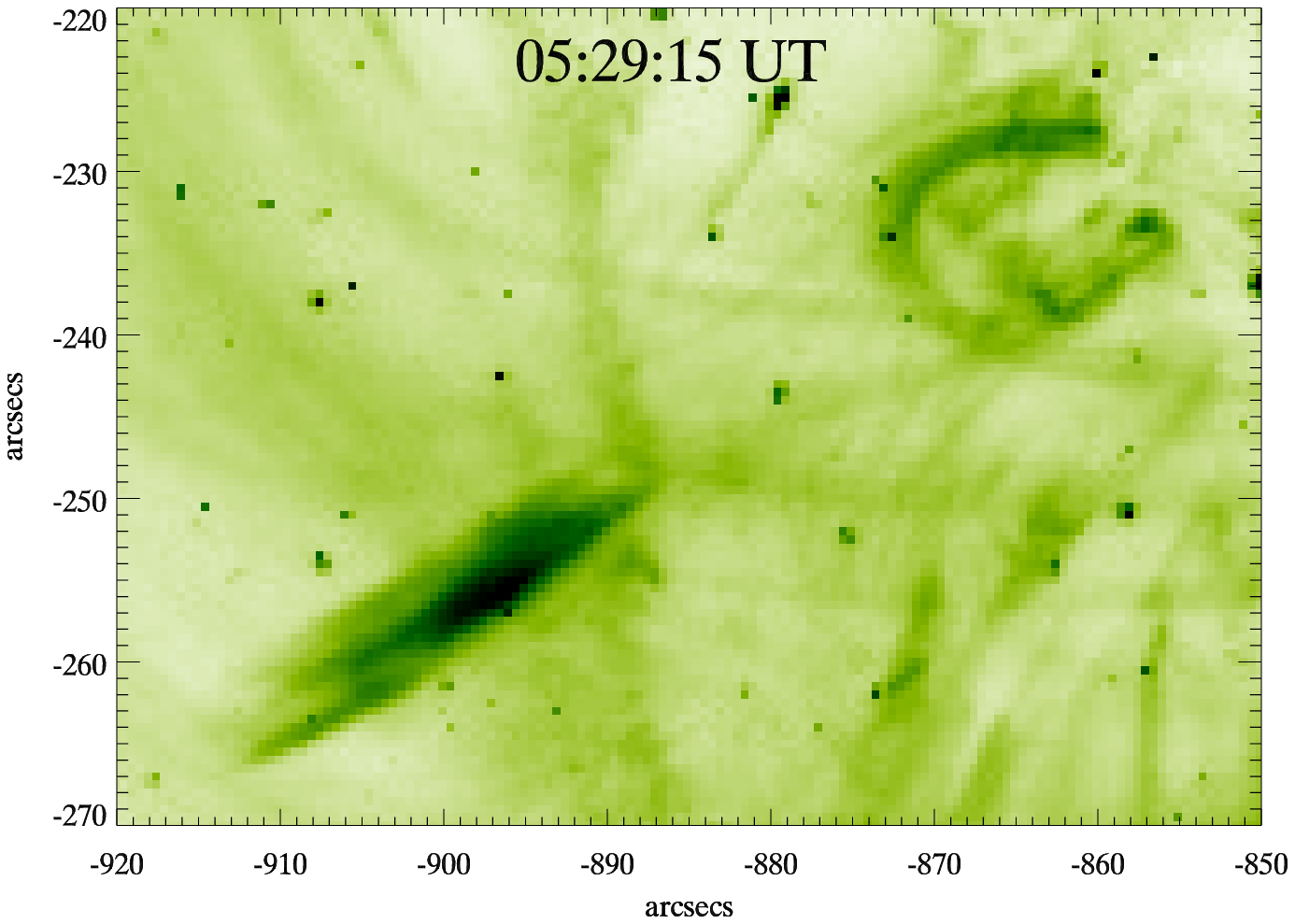}}
\caption{Selected series of TRACE 171 \AA \ images in negative colors showing a moving blob structure above the smaller bright loop
of AR10808.}
\label{fig2}
\end{figure*}

Figure 2 shows the time series of TRACE 171 \AA \ images of the flaring region
in AR10808 during the time interval 05:24-05:29 UT.
The plasma blob (a dark and dense material) is formed between 05:24-05:27 UT above the flare site,
followed by its detachment and propagation upward after 05:27 UT.
TRACE has missed the later part of the observations that is, however, captured by 
H$\alpha$ observations from Culgoora Solar Observatory (cf., Figure 4).
The length and width of the plasma blob, as observed in the
TRACE images in its dynamic phase, are approximately as
22 Mm and 5 Mm, respectively. 
The blob is co-spatially and co-temporally visible in the TRACE 171 \AA\
and H$_{\alpha}$ (6563 \AA ) emissions, and finally vanishes around 05:33 UT.
In the TRACE snapshot at 05:27:55 UT, the tail of the eruptive plasma  
was rooted at (X$_{1}$, Y$_{1}$)$\approx$(-867$"$, -236$"$), while at 05:28:36 UT it reaches 
at (X$_{2}$, Y$_{2}$)$\approx$(-880$"$, -247$"$). Therefore, the  projected distance d$_{12}$ covered 
in $\Delta$t$_{12}$$\approx$41 s is $\approx$12,346 km. Thus, the approximate apparent speed ($v_{1}$) derived 
from these two snapshots is $\approx$ 300 km s$^{-1}$. At 05:29:15 UT, the tail of the blob that is almost constant in shape
for around 3-4 mins, reaches at (X$_{3}$, Y$_{3}$)$\approx$(-888$"$, -250$"$). Therefore, the distance d$_{23}$ covered 
in $\Delta$t$_{23}$$\approx$39 s is $\approx$6200 km. Assuming the 1$"$=725 km scale for the conversion,
therefore, derived approximate speed ($v_{2}$)
is $\approx$ 160 km s$^{-1}$ that can be obtained as the blob propagates. 
Therefore, the average speed of travel is v$_{av}$=(v$_{1}$+v$_{2}$)/2$\approx$230 
km s$^{-1}$. After this the blob quickly vanishes in the corona.
Note that the measured speeds v$_{1}$ and v$_{2}$ are projected speeds of the blob by tracing its detached tail, and the actual speeds may be higher compared to the estimated speeds. Also worth illustrating that these speeds are anyway higher when compared to the local sound-speed ($\sim$150 km s$^{-1}$).
Next, we trace the approximate position of the core of the moving feature in its dynamic phase to
find its projected speed as $\approx$ 175 km s$^{-1}$. Since it is very difficult and somewhat 
superfluous also to track the diffused and faint
head of the moving feature, therefore, we only take the references of the core and tail of blob observed in the TRACE images to estimate
the approximate lower bound speeds. 
In conclusion, the blob plasma moves almost collectively after its origin and detachment from the flaring site
due to reconnection. In the TRACE snapshots during 05:24 UT-05:27 UT, it is clear that there is some interaction 
of the rising plasma column that has detached later on in form of a blob, with the pre-existing fields. The magnetic
field configuration becomes simpler (cf., 05:28-05:29 UT snapshots) at the origin site of the blob after
its detachment. 

\begin{figure}
\centering
\includegraphics[width=10.0cm]{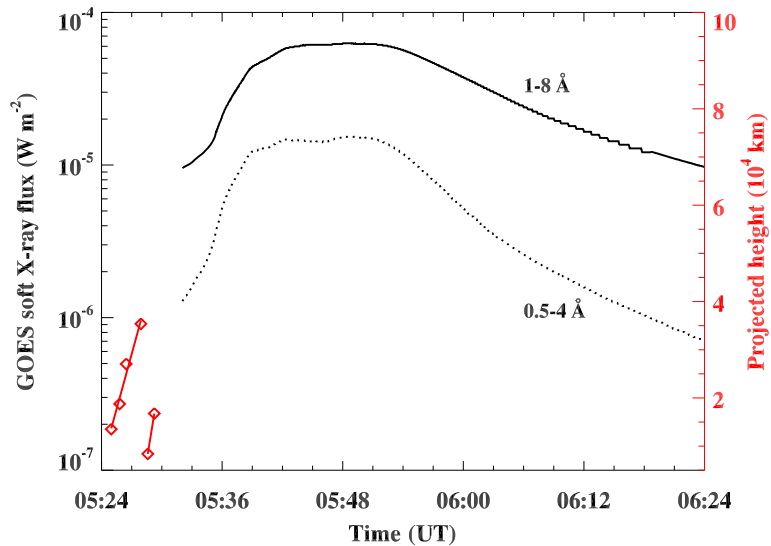}
\caption{The GOES X-ray profiles show a long duration M-class flare, and its 
temporal relation with the rising of plasma blob. The initiation and dynamic 
phases of rise-up are linearly fitted to derive the projected lower bound speeds.
The upper and lower linear fits represent, 
respectively, the initiation and dynamical phase of blob under
the observational baseline of TRACE data.
The linear fit between 05:24-05:27 UT
corresponds to the position of the leading edge of the 
blob while the fit at 05:28-05:30 UT shows the position of its trailing edge.}
\label{fig3}
\end{figure}

Next, the Figure 3 shows the GOES soft X-ray evolution of the M6.2 class solar flare observed 
in the active region and the corresponding projected time-distance plot of the observed plasma
blob in its initiation and dynamic phase during 05:24-05:35 UT is also co-plotted. This is very clear that
the blob rises in a plasma column with a rather low subsonic speed in its build-up phase
(cf., upper linear fit in Figure 3), however, it
accelerates in the dynamic detached phase to a speed of 215 km s$^{-1}$ that crosses the sonic speed level
and propagation becomes supersonic (cf., lower linear fit in Figure 3).
We found that, meanwhile, two more eruptions have been occurred, respectively, at 05:15 UT and 05:50 UT from the same location
(see the attached movie trace-eruptions.avi), however, they are exclusively different from the
observed unique plasma blob. Firstly, other eruptions do not propagate in a definite shape of a blob, and secondly 
they are not visible simultaneously in the cool H$\alpha$ temperature. 
However, the multiple supersonic eruptions from the same location may be the 
collective signature of repetitive reconnections ongoing at the flare site. 
Figure 3
also depicts the linear fit in the projected distance of the travel of blob 
as measured from the TRACE image sequence, and it is also given both in its initiation and detached dynamic phases.
The linear fit between 05:24-05:27 UT
corresponds to the position of the leading edge of the 
blob in the initiation phase, while the fit at 05:28-05:30 UT shows the position of its trailing edge in the dynamical phase. 
The speed measured in the dynamical phase matches well
with the average speed of 230 km s$^{-1}$ as measured crudely by analyzing TRACE image sequence. We note that
the plasma column was built well before the flare event, while the dynamic
phase of the blob is achieved just before the rising phase of the flare. The rising speed
of the plasma column before the detachment of the blob
is $\approx$126 km s$^{-1}$ before the occurence of the flare event. This 
phase of the rising of plasma column is clearly evident in Figure 2 during
05:24-05:28 UT. The denser and brighter material is also filled
in the plasma column during this time span, which
later detached in form of the observed plasma blob. We measure the speed
of the rise-up 
of this plasma column in the initiation phase 
before the detachment of blob by tracing 
its leading edge during the course of time (cf., snapshots during 
05:24-05:28 UT in Figure 2, 
and upper linear fit in Figure 3).
While in the dynamic phase, the
blob gains a supersonic speed of $\approx$215 km s$^{-1}$ as per the linear fit measurements, 
during early rising phase of the gradual M-class solar flare. This measured speed
is now much more accurate measurement. The average sound speed at the Fe XI 171 \AA\ formation temperature (T$_{f}$ = 1$\times$10$^{6}$ K) is $\sim$150 km s$^{-1}$. Therefore, the blob propagates with a supersonic speed, quickly changes its shape, and finally becomes fainter against the background by most likely dissipating its energy and material draining back to the low atmosphere (see the movie trace-eruptions.avi).
It should be noted that the whole process has two phases: first the initiation phase in which the plasma column and associated cylindrical plasma  blob is build-up, followed by the second dynamical stage when the plasma blob become detached and moves up in the corona
with a supersonic speed within few minute time scale. The true speed (projected) what is quoted here is in the dynamical phase and is consistent both with H$_{\alpha}$ and TRACE observations.

The H$_{\alpha}$ observations of this active region was carried out at Culgoora Solar Observatory at Narrabri, Australia by using a 12 cm f/15 Razdow solar patrol telescope equipped with an H$_{\alpha}$ filter of the Lyot type that has a pass bandwidth of 0.5 \AA\ .  The raw images were recorded by an 8 bit 1024$\times$1024 pixels CCD camera system with pixel size 6.7$\times$6.7 $\mu$m$^{2}$. The spatial resolution of the camera is 2$^{\prime\prime}$ per pixel. The typical cadence for the present observations is at least 1 frame per minute. The observations are carried out with a FOV of the full-disk solar image.
Figure 4 shows a sequence of data of the propagating blob emphasized by a yellow arrow from AR10808 before an M-class flare during 
05:26 UT--05:36 UT. The difference images of Figure 4 are a 160$^{\prime\prime}$$\times$100$^{\prime\prime}$ partial field 
of view provided to us by Culgoora Solar Observatory under its data use policy. Although, the resolution of the images
is not very high, it should be noted that this is the only H$\alpha$ observations that are available from the ground as this unique
event occurred in its day-light time zone. The temporal H$\alpha$ observations are thus very useful and show the unique signature of the propagation of a {\it low-temperature} counterpart of the plasma blob observed by TRACE propagating away from the flare site towards the eastern limb in a projection. 
A careful investigation of the H$_{\alpha}$ image sequence shows that the blob is initially of a cylindrical shape of the length of $\approx$22 Mm and width of $\approx$5 Mm. The length and width are found to be consistent with the observed TRACE images in the dynamic phase of the blob. Therefore, we conclude that initially the length to width ratio is 4.4. 
It should also be noted that not only the position but also the length to width ratio may also be influenced 
by projection effects. The blob changes its shape quickly and is faded within 4-5 minutes of its evolution as visible in H$_{\alpha}$ at 05:28 UT. The measured lower bound speed of this propagating blob is found to be $\sim$200 km s$^{-1}$. 
This is again the projected speed of the blob, and the actual speed may be higher compared to this lower bound estimated speed. The average sound speed of the H$_{\alpha}$ formation temperature (T$_{f}$ = 1$\times$10$^{4}$ K) is 15 km s$^{-1}$. 
Therefore, the blob propagates with supersonic speed.

\section{Theoretical Modelling and Interpretation}
\label{obs}
As we have explained in the previous section, the analyses of the H$_{\alpha}$ and TRACE 171 \AA\ temporal data between
05:24 UT--05:36 UT well before an M6.2 class flare from AR10808 
on 9 September 2005 show the observational evidence of the supersonic
plasma blob that propagates in the higher solar atmosphere
with a supersonic speed of $\sim$ 215 km s$^{-1}$. 
\begin{figure*}
\centerline{
\hspace*{-0.04cm}
\includegraphics[width=5cm]{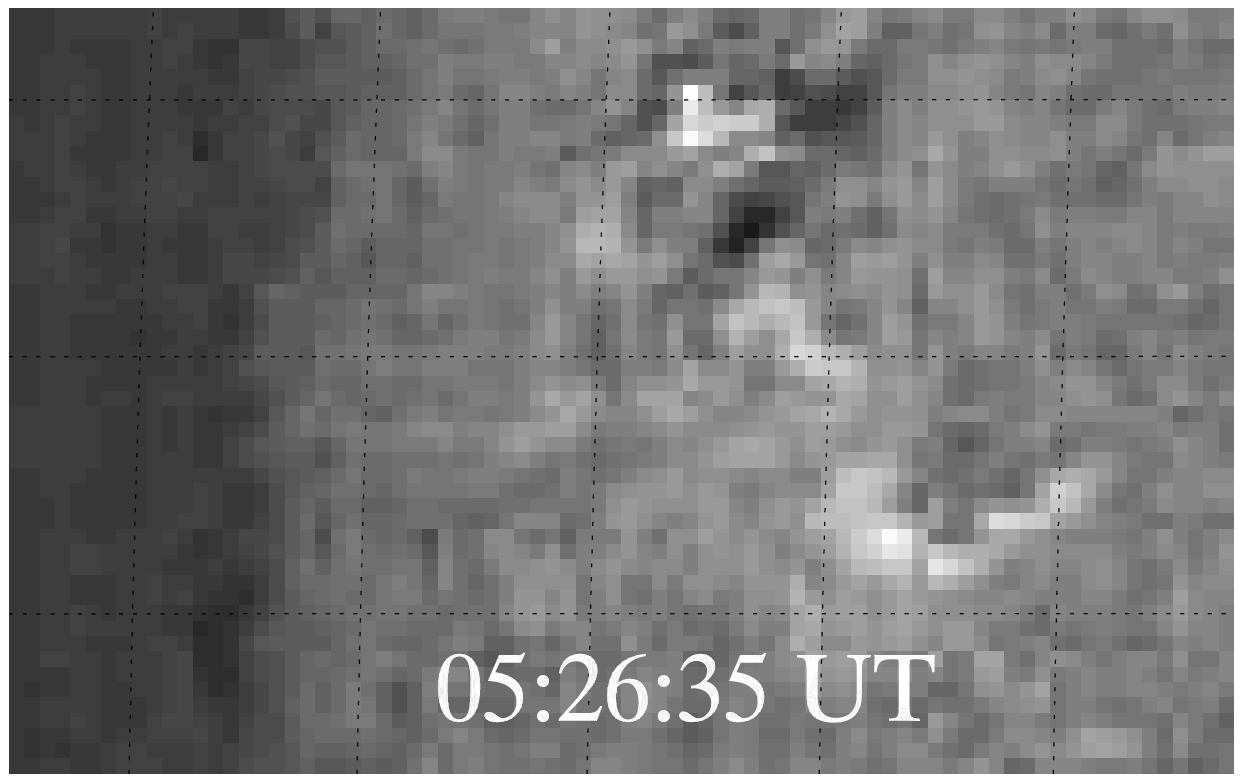}
\includegraphics[width=5cm]{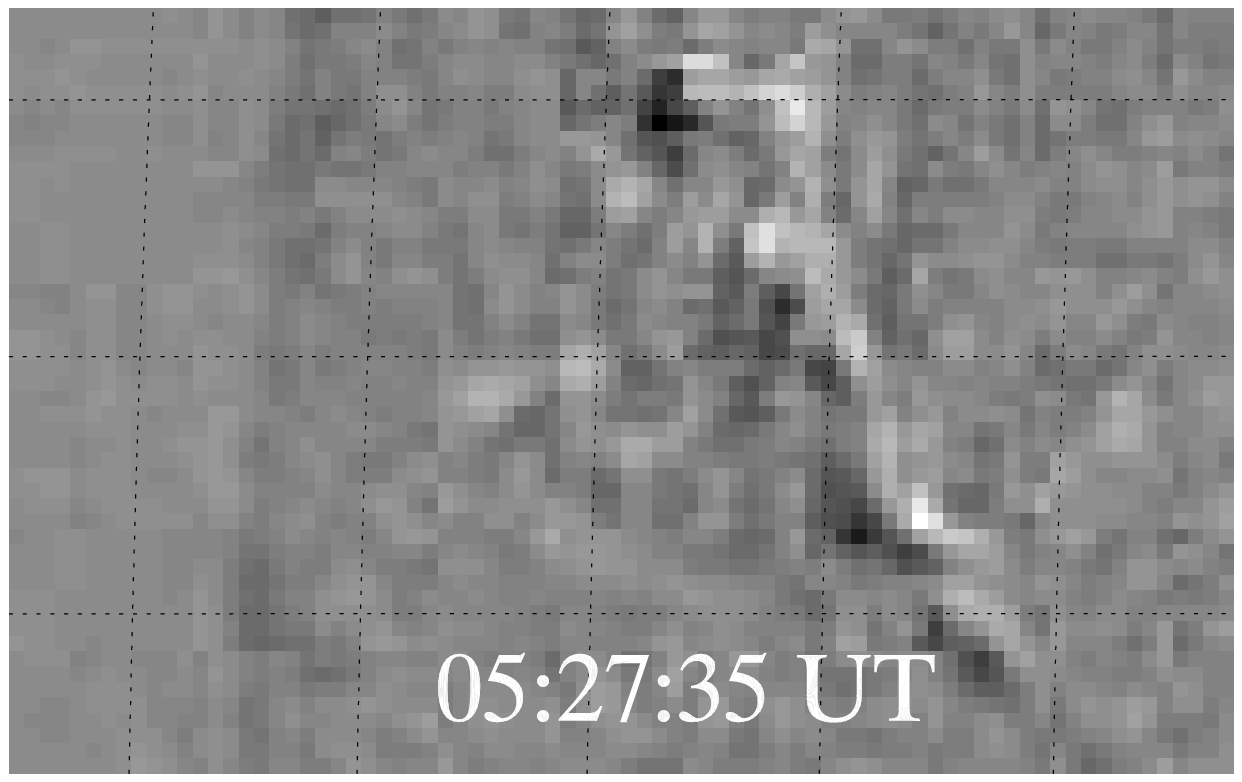}
\includegraphics[width=5cm]{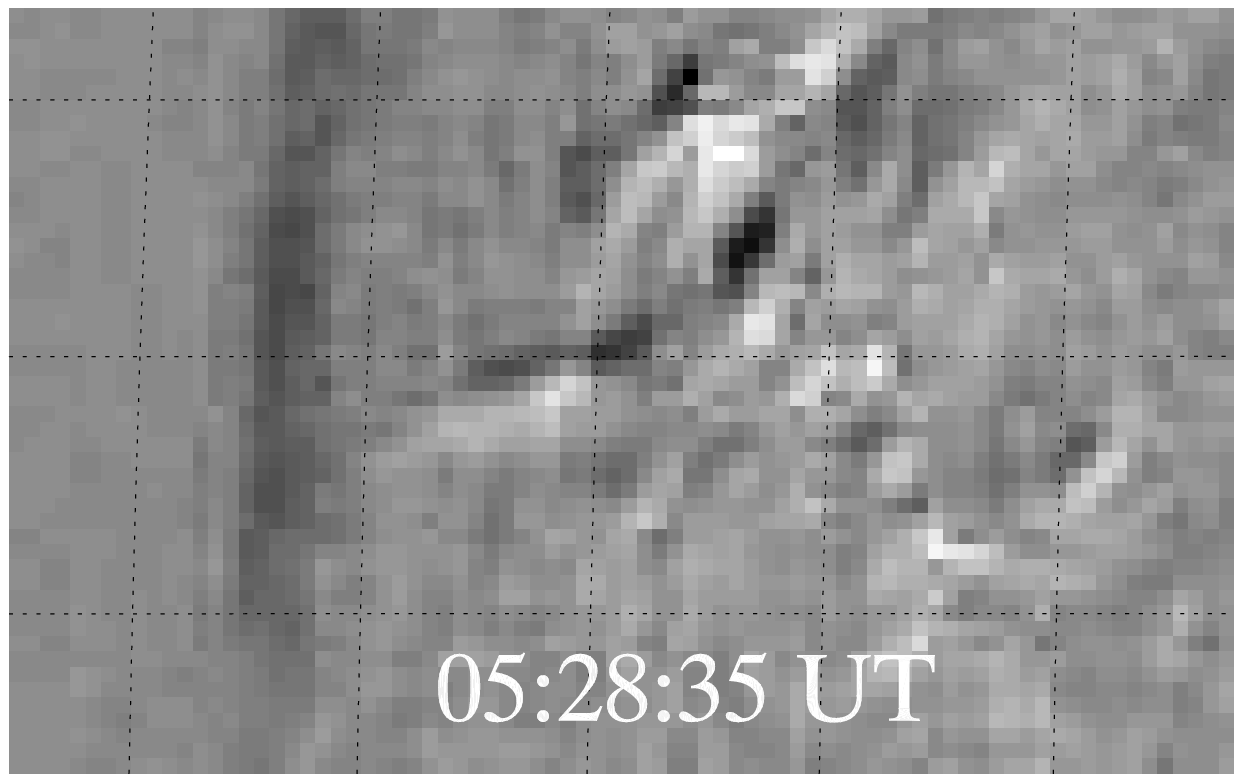}
\thicklines
$\color{yellow} \put(-120,42){\vector(1,0){15}}$
}
\centerline{
\hspace*{-0.08cm}
\includegraphics[width=5cm]{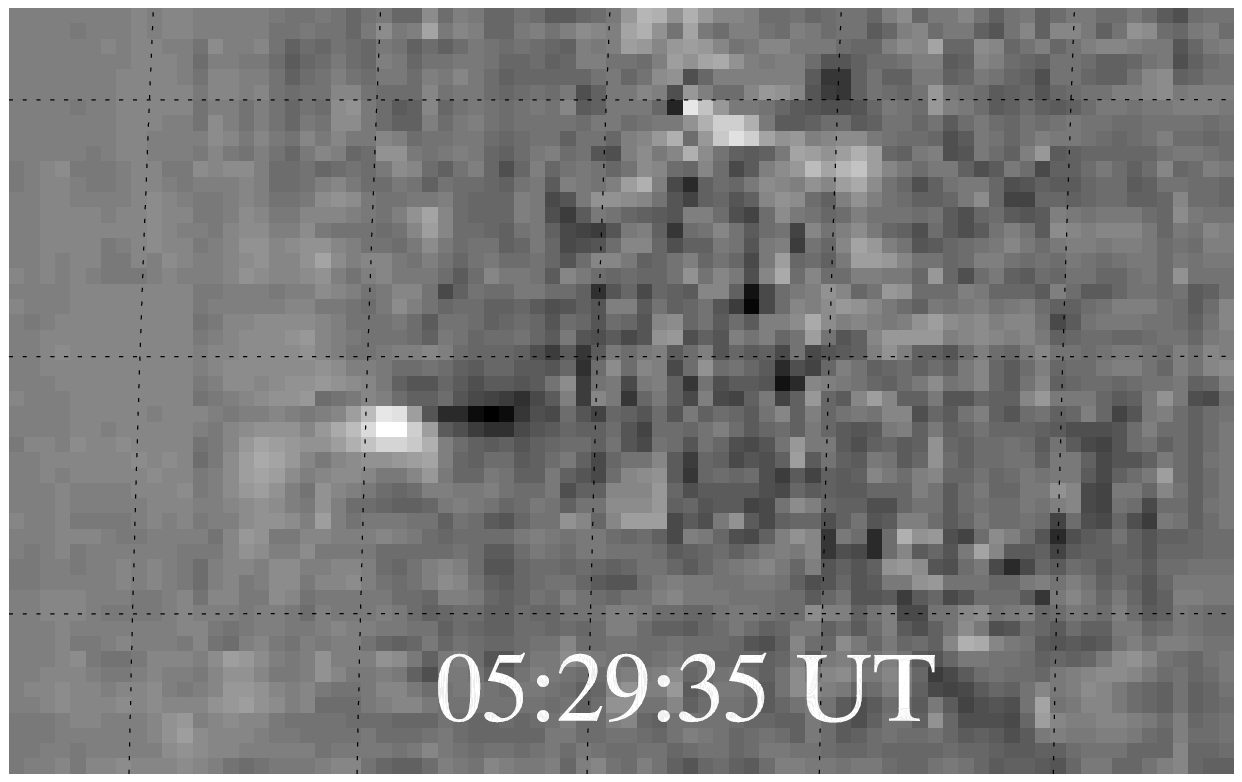}
\hspace*{-0.28cm}
\thicklines
$ \color{yellow} \put(-122,42){\vector(1,0){15}}$
\includegraphics[width=5cm]{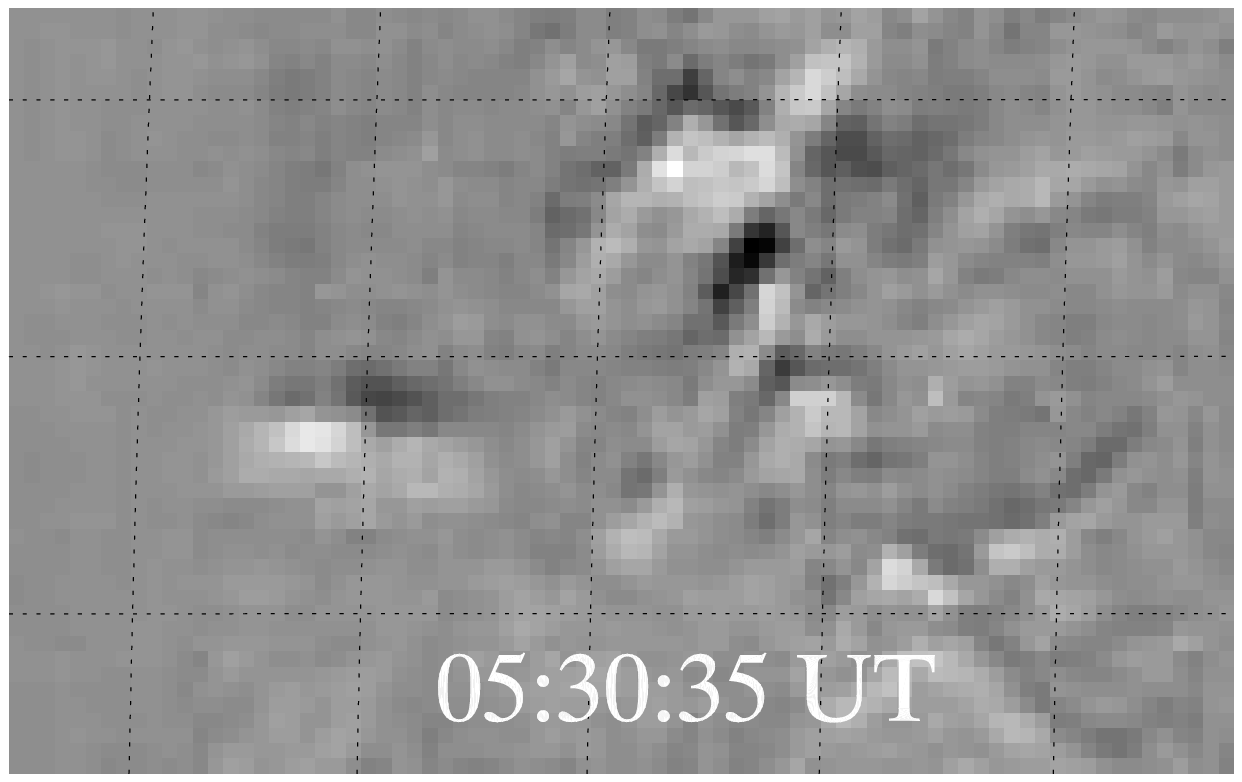}
\hspace*{-0.28cm}
\thicklines
$ \color{yellow} \put(-135,41){\vector(1,0){15}}$
\includegraphics[width=5cm]{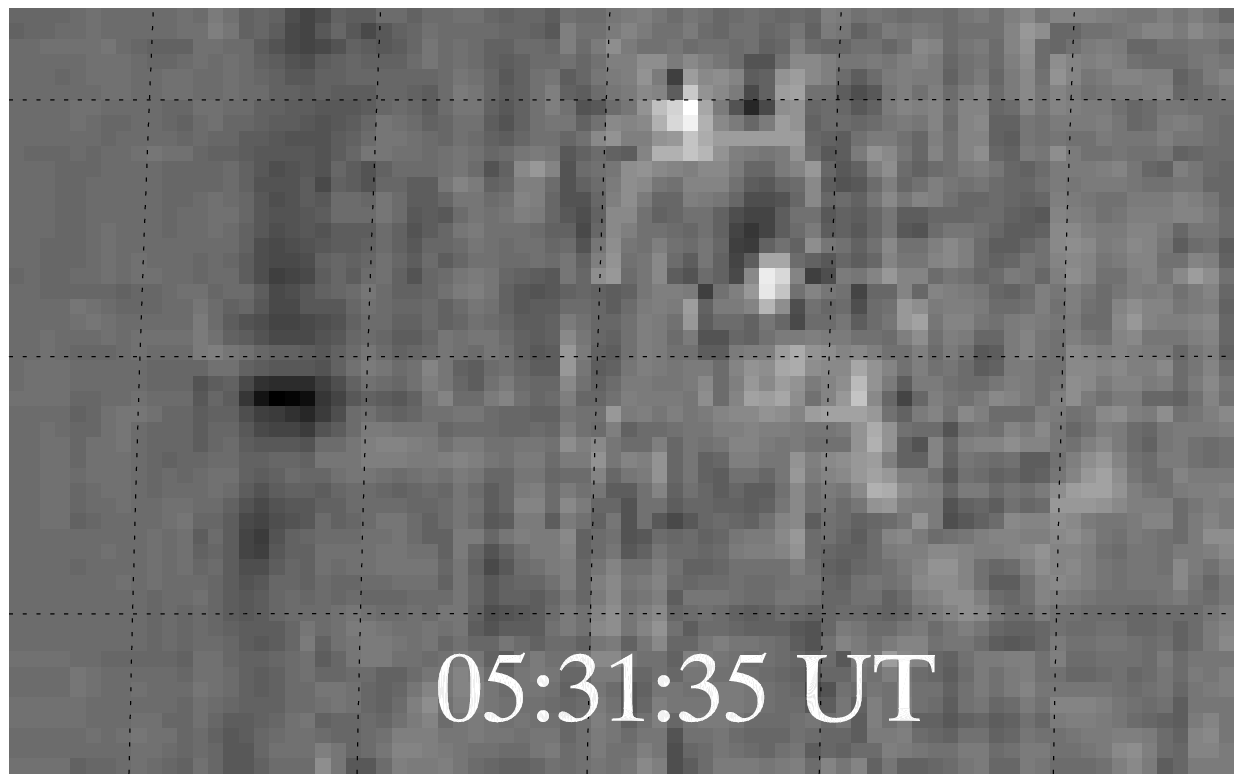}
\thicklines
$ \color{yellow} \put(-137,44){\vector(1,0){15}}$
}
\centerline{
\hspace*{-0.2cm}
\includegraphics[width=5cm]{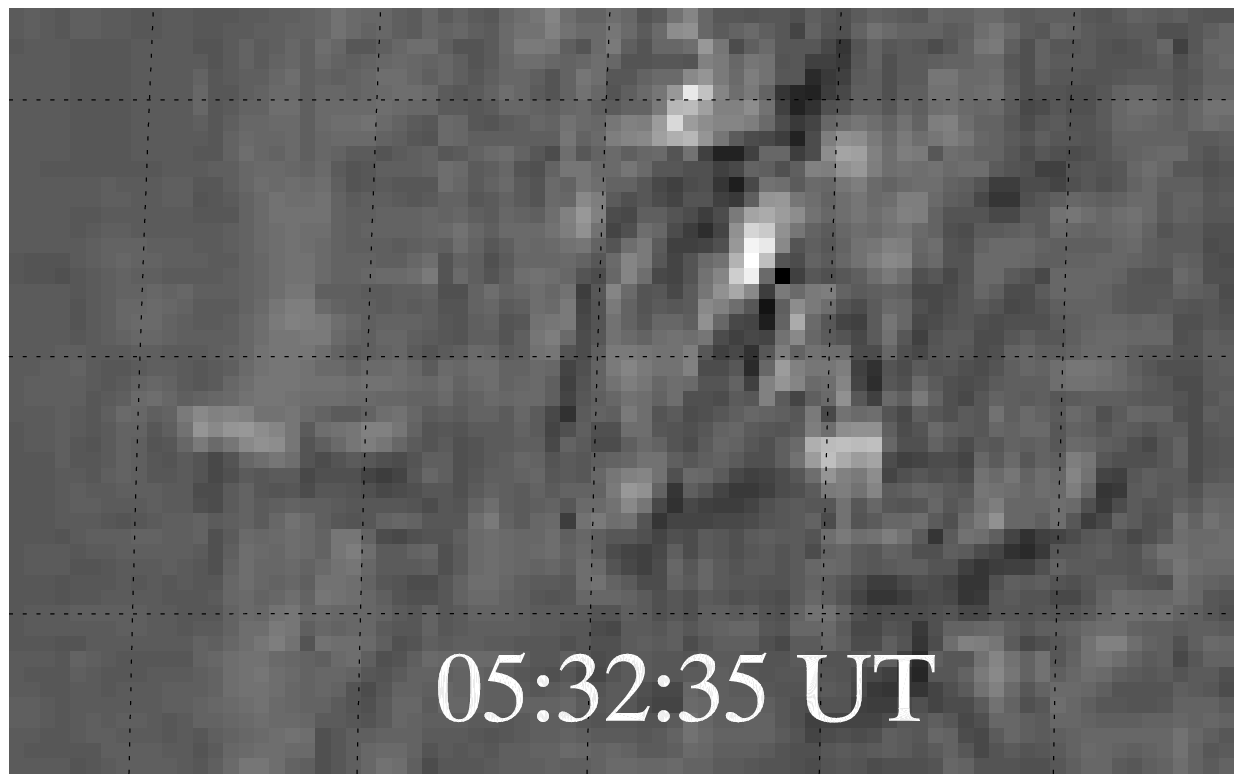}
\includegraphics[width=5cm]{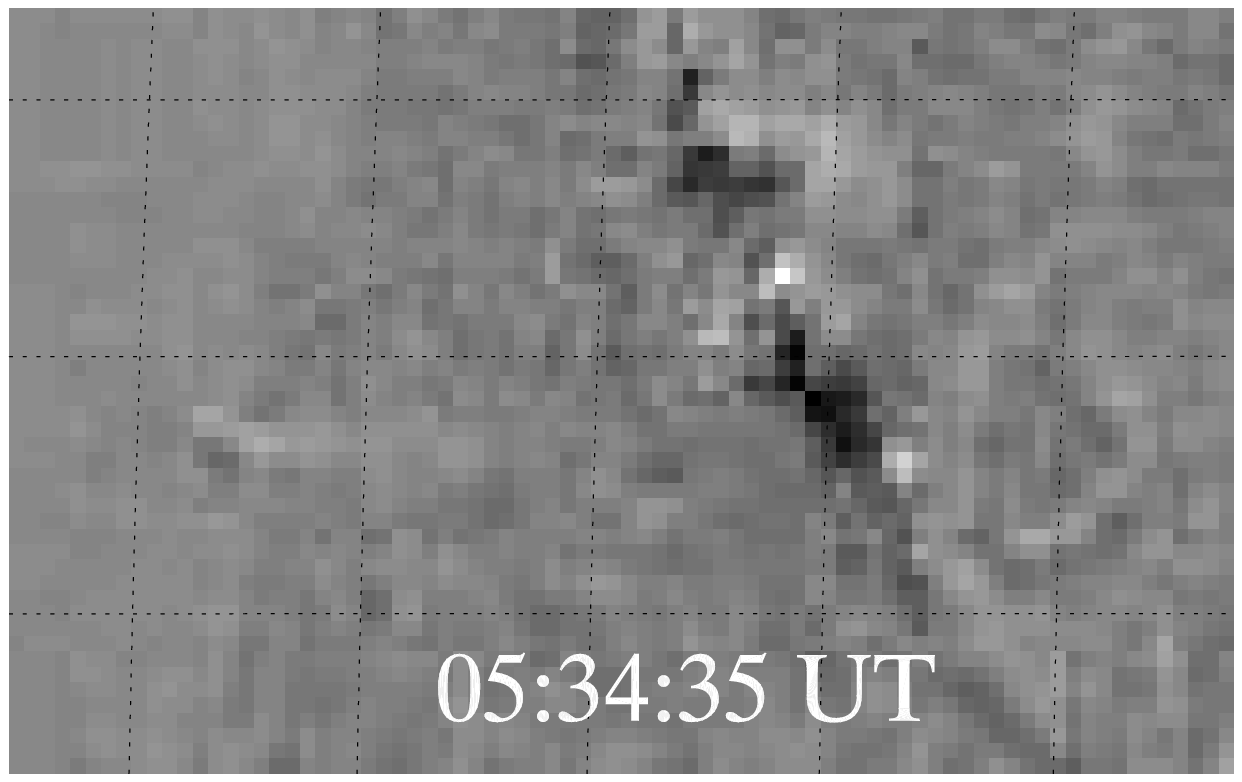}
\includegraphics[width=5cm]{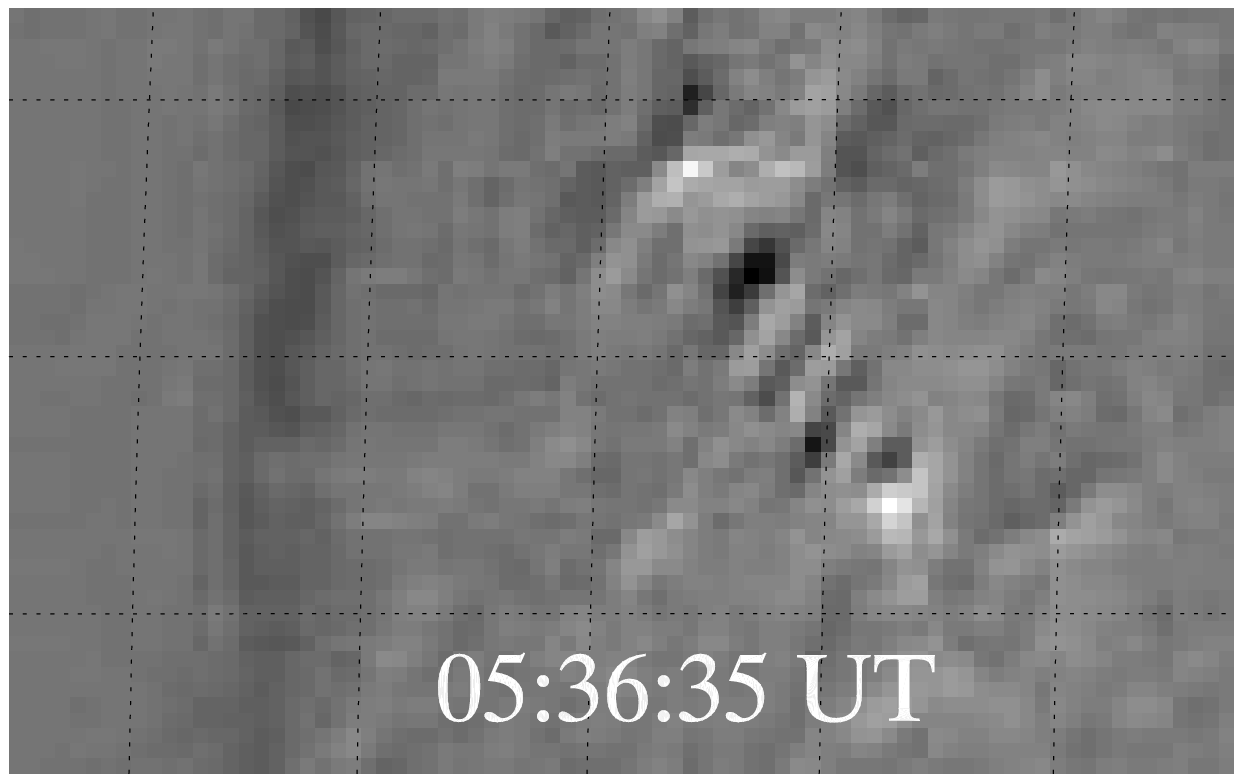}
}
\caption{Selected H$_{\alpha}$ difference images obtained from H$_{\alpha}$ observations of the Culgoora Solar Observatory showing a moving blob structure towards the eastern limb highlighted by yellow arrow. The size of each image is 160$^{\prime\prime}$$\times$100$^{\prime\prime}$.}
\label{fig4}
\end{figure*}
We firstly explore the possibility of the generation of this supersonic blob 
as a reconnection-generated plasmoid. The observed blob may be the 
plasmoid as recorded previously by \inlinecite{Mano03} due
to magnetic reconnection during the rising phase of an M-class flare. 
They have observed the motion of plasma blobs associated with a sigmoid, Moreton wave,
CME injection in the interplanetary space, and associated radio bursts
from AR9393 on 2 April 2001 at 11:00 UT. They have also found magnetic
reconnection around a coronal null point as a cause of such eruptions and 
associated phenomena. Recently, the multiple plasmoid and its dynamics have also been 
observed in solar flares \cite{barta2008,nishizuka2010}
. However, such 
plasmoids are the high-speed plasma blobs that accelerate outwards
the reconnection point in the solar atmosphere and usually are associated with 
the rising phase of the 
solar flares. These plasmoids are often released with CME ejections, supersonic
down-flows that causes the HXR emissions, and radio bursts \cite{Mano03}.
In the present case, the supersonic blob is observed
well {\it before} the M 6.2 class flare, and there is no observational
evidence of associated CMEs and radio bursts even in this most
violent super-active region associated with this flare on 
9 September 2009. \inlinecite{kundu2001} have reported that the plasmoid ejecta are associated with
metric/decimetric emission that starts significantly after
the impulsive hard X-ray (20 keV) and microwave bursts.
In this event, however,  we do not observe any metric/decimetric or microwave
radio emissions as evident in the Culgoora and Learmonth radio spectra, and no
impulsive hard X-ray emissions are evident at that time as per RHESSI X-ray flux profiles.
These observed facts possibly exclude the formation 
and acceleration of the supersonic plasmoid excitation in our case. 
In conclusions, although the morphology of the 
observed blob and its dynamics are not favorable for the plasmoid 
ejection, sometime it may be difficult to distinguish between
the plasma blob formed by the 
reconnection rate change and real plasmoids \cite{Bat07}.
The detailed analyses of the magnetic field topology 
can only differenciate between these two.
Moreover, a single plasmoid does not need to be accompanied with radio/HXR
emissions always. The most likely interaction of multiple plasmoids may be responsible for
particle acceleration and related emissions.

The observed plasma blob may be a reconnection-generated jet as per its typical length, time,
velocity scales of the coronal jets \cite{nistico2009}.
However, we suggest that the observational evidences point towards more
in the form of a pulse driven blob, which may be a special type
of detached jet rather than a classical jet-like structure. Secondly, the blob
is visible both in H-alpha and TRACE 171 \AA\ with similar shape and dynamics, which
indicate that it is formed by {\it multi-temperature} plasma.
The bi-directional EUV jets \cite{Innes97} are the natural
consequences of the flow induced coronal X-type reconnections
as theorized by many workers \cite{Pets64,roussev2001a}. However, the various types of 
the classical EUV jet as well as such type of observed plasma
packets that move radially outwards from the lower solar 
atmosphere, can either be a direct consequence of the low-atmospheric 
reconnection of the low-lying loop systems \cite{nistico2009}
or the energy build-up due to some magnetic instabilities \cite{Pas08}.
Therefore, the observed blob may also be the consequence of the
low-atmospheric reconnection as also evident in the observations.

If the supersonic blob is not a typical form of the ejecta of flaring active regions, e.g.,
reconnection-generated classical jets, plasmoids etc, then the question arises about its identification and the 
most probable
drivers during its initiation and dynamic phases. Recently
\inlinecite{zaqarashvili2010} have reported the propagation of sausage solitons 
in the quiet Sun chromosphere using Hinode/SOT observations. They have reported the 
propagation of a supersonic plasma blob in the solar chromosphere, which retain the
almost constant shape (length to width ratio) and obeys the soliton solution. 
The propagating blob reported in this paper here also moves with a supersonic speed, but changes its relative 
intensity amplitude as well as the shape (length to width ratio) within 3 
minutes of its launch at 05:27 UT and, then, quickly vanishes against the background plasma.
Therefore, the soliton description is not the most evident and plausible explanation
associated with the observations reported here.
\begin{figure}
\centering
\includegraphics[width=10.0cm]{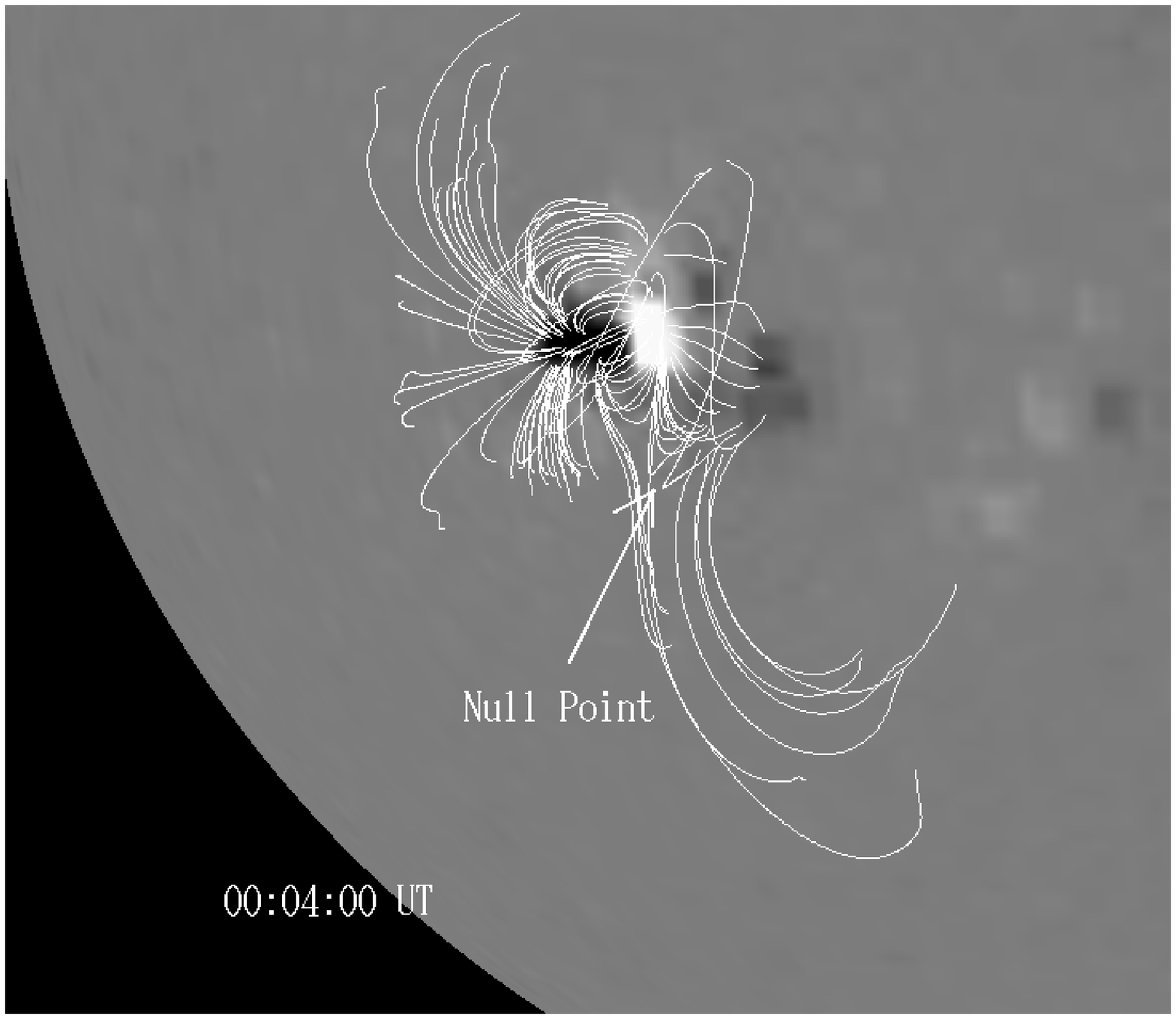}
\caption{Potential Field Source
Surface (PFSS) extrapolation overplotted on the MDI map of 00:04 UT at 9 September 2005
shows the configuration of overlying loops and the position of magnetic null point.
}
\label{fig1}
\end{figure}

The interpretation of a kink pulse/wave may be another possibility to drive such observed plasma blob/jet
in its dynamic phase \cite{Zaqa07,Cirt07}. However,
the axial transversal displacements or a curved form of the pulse is not evident 
in the observations. The density and thus intensity enhancement in the associated 
inclined flux tube is also not evident as reported by \inlinecite{cooper2003}. Therefore, the
possibility of the kink pulse as a driver of the supersonic blob in its
dynamic phase, is also out of the scope under the baseline 
of the observations.
Another and perhaps the most plausible possibility is then the excitation of a
fast MHD pulses/wave train in the flaring active region as a manifestation of the 
supersonic blob in its dynamic phase. The blob propagates 
with supersonic speed and vanishes quickly in the corona within
3-4 minutes. The blob may be energized by the fast 
wave trains generated in the solar atmosphere \cite{Nak04}.
In such a case, we should observe the quasi-periodic
multiple rise and fall of the blob material, which
is indeed not evident in the observations presented in the paper.
Therefore, the fast MHD pulse train is an unlikely driver for the
observed supersonic plasma blob made-up by the multi-temperature 
plasma. In conclusion, the observed blob is 
neither some classical phenomena of the solar atmosphere
(e.g., Y-shaped typical jets, plasmoids, up-flows etc),
nor it is the signature of the peculiar MHD phenomena
(e.g., sausage solitons, fast MHD wave trains,
kink pulses etc).

Therefore, we suggest that the reconnection-generated velocity pulse
at flaring site may be the most likely driver of the plasma blob.
The energy release at reconnection site in the lower
solar atmosphere perturbs the plasma velocity and most
possibly creates a velocity pulse that trigger the plasma eruption 
in the form of a supersonic blob.
To the best of our knowledge, this is the first observational signature of the excitation
of a velocity pulse driven supersonic plasma blob in the solar atmosphere
above an active region in the vicinity of a flare site.
The most likely possibility is that the low-lying magnetic loops reconnect
with each other. This
reconnection is at its beginning in the steady state and most
probably with a single
dissipation region. Since, this reconnection region has so far not reached 
its dynamical state with many
plasmoids created, therefore, low number of particles is accelerated and we do not
observe either radio or HXR emission. Hence, this steady
reconnection gives only the rise of the plasma jet which we observe from the very
beginning (e.g., Figure 2, 5:24 UT snapshot). The formation
of the plasma jet during steady state reconnection further changes into the
detached blob propagating in corona. As the reconnection
rate and reconnection regime change, a velocity pulse may be generated
to accelerate the plasma blob. Due to the
mass conservation, a density blob is formed in the outflow jet, which detaches
from the reconnecting loops and propagates upwards until it vanishes 
in the solar corona.

We also explore the details of magnetic field topology 
to understand the most possible generation of such plasma 
dynamics that we have been observed.
The magnetic field configuration and
topological distribution of its positive and negative polarities are shown in Figure 1.
It is clear from Figure 1 that the positive polarity region 
is stretched towards the west with a remote tongue, while the
negative polarity is stretched towards the east with a remote tongue
in the opposite direction as an indicator of the emergence of the twisted magnetic fluxes
(e.g., Archontis and Hood, 2010; Li et al., 2007).
The Potential Field Source Surface (PFSS) extrapolation (Figure 5) of the SoHO/MDI at 00:04 UT on 09
September 2005 also reveals the formation 
of a coronal null point above a positive polarity sunspot 
which is joining the two segments of oppositely lying
negative polarities by bipolar loop systems. This is a very preliminary
configuration of the AR magnetic field well before the event, and later
phase was also associated with the high flux-emergence and build-up
of the complexity in the region (cf., Figure 1, right-panel). This configuration might
force the steady state magnetic reconnection in the lower part of the atmosphere.
The PFSS is slightly shifted in its centroids so that we see clearly 
the extrapolated field and its configuration more towards our line-of-sight.
In this parasitic region, the central positive 
polarity spot is a unipolar flux region and connecting
with the two opposite polarity regions on both sides.
This is similar situation as the formation of a dome-like
fan surface where the symmetry axis is known as spine containing
a null point at the intersection with the fan dome \cite{asc2004}.
Our observations (Figure 5) show a clear evidence of the
formation of a parasitic region, i.e., the fan dome, null-point and spine
at the flare site in the active region.
However, the energy release may be taking place well below this magnetic
null point due to pre-emergence and therefore the reconnection process with 
the existing field lines in the lower solar atmosphere. The field topology in Figure 5 is following the
standard magnetic topology of a 3-D reconnection process via 
a separator dome (see Figure 10.27 of Aschwanden, 2004 and Fletcher et al., 2001).
The PFSS extrapolation schematic diagram (extrapolated field lines
over SoHO/MDI image) approximately matches well with the standard scenario of the 3-D reconnection
and thus the re-arrangements of the field lines and mass motion along
open field lines in the vicinity of the null point.  However, it seems that
the steady state reconnection generates the plasma jet in the lower solar atmosphere.
There may be some small energy release also during the magnetic field rearrangements 
of the dynamical reconnection process, however, it is clear that there is no bulk
heating of the plasma before achieving the flare maximum. Moreover, there is no 
expansion or diffusion of the detached plasma along the field lines outward.
Therefore, we discard the acceleration of the blob due to the generation
of a thermal pulse by bulk heating.
This magnetic field re-arrangements during the reconnection process 
is found to be the driver of a velocity pulse along the radial direction outward and the observed driven supersonic 
plasma blob. This type of reconnection scenario drives the plasma
outward in the radial direction along the spine-lines due to the low-atmospheric
reconnection between the low-lying loop systems. This reconnection generates a velocity
pulse, which associates with a shock steepened in the corona to drive the 
plasma blob. Therefore, we  have the 
consistent observation of an uni-directional detached jet or plasma blob propagating outwards.
Recently, the reconnection-generated velocity 
pulse driven jet \cite{Sri11}, and macro-spicules \cite{Mur11}
have already been modeled. However, this is the first attempt to
numerically model the reconnection generated and velocity pulse driven plasma blob in the solar corona.

Here, the most generic and real situation is that the
plasma blob is launched due an initial radial velocity
pulse along the magnetic field lines at the triggering site of the blob. The blob is 
generated approximately 5-6 Mm above the photosphere with the start of the reconnection
process. The reconnection at this height in the low-lying quadrapolar loop system may generate the velocity pulse
which may cause the formation of the supersonic plasma blob 
propagating up in the corona. In the next
section, we model numerically the propagation
of the supersonic plasma blob generated by a velocity pulse 
at the reconnection site.
%
\subsection{Numerical simulations}
We consider a gravitationally stratified solar atmosphere which is described by 
the ideal 2D magnetohydrodynamic (MHD) equations:
\beqa
\label{eq:MHD_rho}
{{\partial \varrho}\over {\partial t}}+\nabla \cdot (\varrho{\bf V})=0\, , 
\\
\label{eq:MHD_V}
\varrho{{\partial {\bf V}}\over {\partial t}}+ \varrho\left ({\bf V}\cdot \nabla\right ){\bf V} = 
-\nabla p+ \frac{1}{\mu}(\nabla\times{\bf B})\times{\bf B} +\varrho{\bf g}\, ,
\\
\label{eq:MHD_p}
{\partial p\over \partial t} + \nabla\cdot (p{\bf V}) = (1-\gamma)p \nabla \cdot {\bf V}\, , 
\hspace{3mm} 
p = \frac{k_{\rm B}}{m} \varrho T\, , 
\\
\label{eq:MHD_B}
{{\partial {\bf B}}\over {\partial t}}= \nabla \times ({\bf V}\times{\bf B})\, , 
\hspace{3mm} 
\nabla\cdot{\bf B} = 0\, .
\eeqa
Here ${\varrho}$ is the mass density, ${\bf V}$ the flow velocity,
${\bf B}$ the magnetic field, $p$ the kinetic gas pressure, $T$ temperature, 
$\gamma=5/3$ the adiabatic index, ${\bf g}=(0,-g)$ the gravitational acceleration where
$g=274$ m s$^{-2}$, 
$m$ denotes the mean particle mass and $k_{\rm B}$ the Boltzmann's constant.

We assume that the solar atmosphere is in static equilibrium (${\bf V}_{\rm e}=0$) with a force-free magnetic field, 
%
$(\nabla\times{\bf B}_{\rm e})\times{\bf B}_{\rm e} = 0\, $.
%
At this equilibrium the pressure gradient is balanced by the gravity force, 
%
$-\nabla p_{\rm e} + \varrho_{\rm e} {\bf g} = 0\, $.
%
Here the subscript $'$e$'$ corresponds to equilibrium quantities.
Using the ideal gas law and the $y$-component of hydrostatic
pressure balance, 
we express
the equilibrium gas pressure and mass density as
\beqa
\label{eq:pres}
p_{\rm e}(y)=p_{\rm 0}~{\rm exp}\left[ -\int_{y_{\rm r}}^{y}\frac{dy^{'}}{\Lambda (y^{'})} \right]\, ,\hspace{3mm}
\label{eq:eq_rho} 
\varrho_{\rm e} (y)=\frac{p_{\rm e}(y)}{g \Lambda(y)}\, .
\eeqa
Here
%
$\Lambda(y) = k_{\rm B} T_{\rm e}(y)/(mg) $
%
is the kinetic pressure scale-height, and $p_{\rm 0}$ denotes the kinetic gas
pressure at the reference level that we choose in the solar corona at $y_{\rm r}=10$ Mm. 
We adopt 
an 
equilibrium temperature profile {\bf $T_{\rm e}(y)$} for the solar atmosphere 
that is close to the VAL-IIIC atmospheric model of \inlinecite{Ver1981}.

We assume that the initial magnetic field satisfies a current-free condition, 
$\nabla \times \vec B_{\rm e}=0$, and it is specified by the magnetic flux function, $A$, 
such that
$
\vec B_{\rm e}=\nabla \times (A\hat {\bf z})\, .
$
%
We set an arcade magnetic field by choosing 
%
\begin{equation}
A(x,y) = B_{\rm 0}{\Lambda}_{\rm B}\cos{(x/{\Lambda}_{\rm B})} {\rm exp}[-(y-y_{\rm r})/{\Lambda}_{\rm B}]\, .
\end{equation}
$B_{\rm 0}$ is the magnetic field at $y=y_{\rm r}$, and the magnetic scale-height is
%
${\Lambda}_{\rm B}=2L/\pi\, $.
%
We use $L=30$ Mm. 

We initially perturb 
the above equilibrium impulsively by a Gaussian pulse in the 
vertical component of velocity, $V_{\rm y}$, viz., 
\beq\label{eq:perturb}
V_{\rm y}(x,y,t=0) = A_{\rm v} \exp\left[ -\frac{(x-x_{\rm 0})^2+(y-y_{\rm 0})^2}{w_{\rm p}^2} \right]\, .
\eeq
Here $A_{\rm v}$ is the amplitude of the pulse, $(x_{\rm 0},y_{\rm 0})$ is its initial position and
$w_{\rm p}$ denotes its width. We 
choose and hold fixed 
$x_{\rm 0}=4$ Mm, {\bf $y_{\rm 0}=5$} Mm, $w=2$ Mm, and $A_{\rm v}=300$ km s$^{-1}$. 

\begin{figure*}
\centering
\mbox{
\includegraphics[width=5.5cm,height=8.5cm, angle=90]{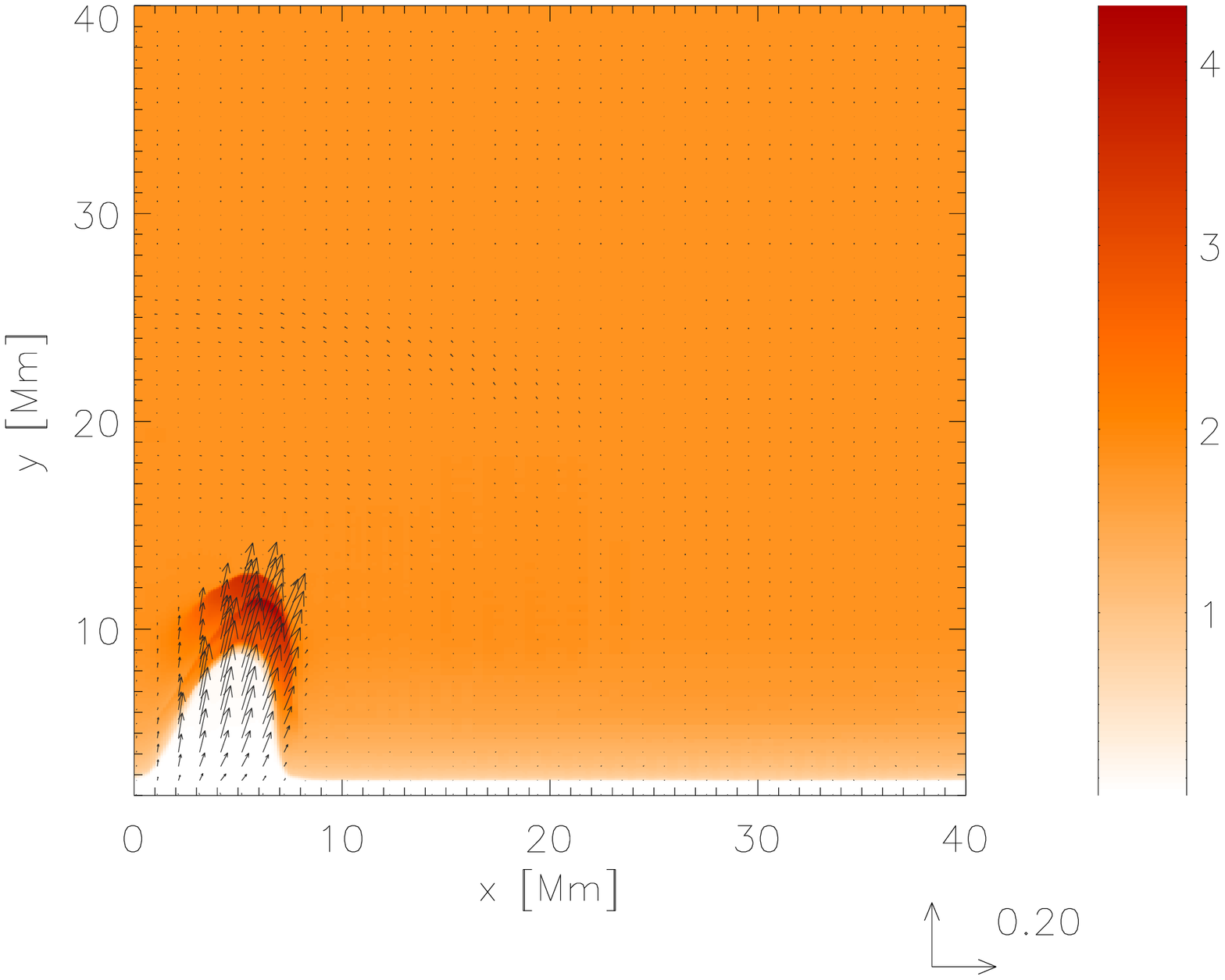}
\includegraphics[width=5.5cm,height=8.5cm, angle=90]{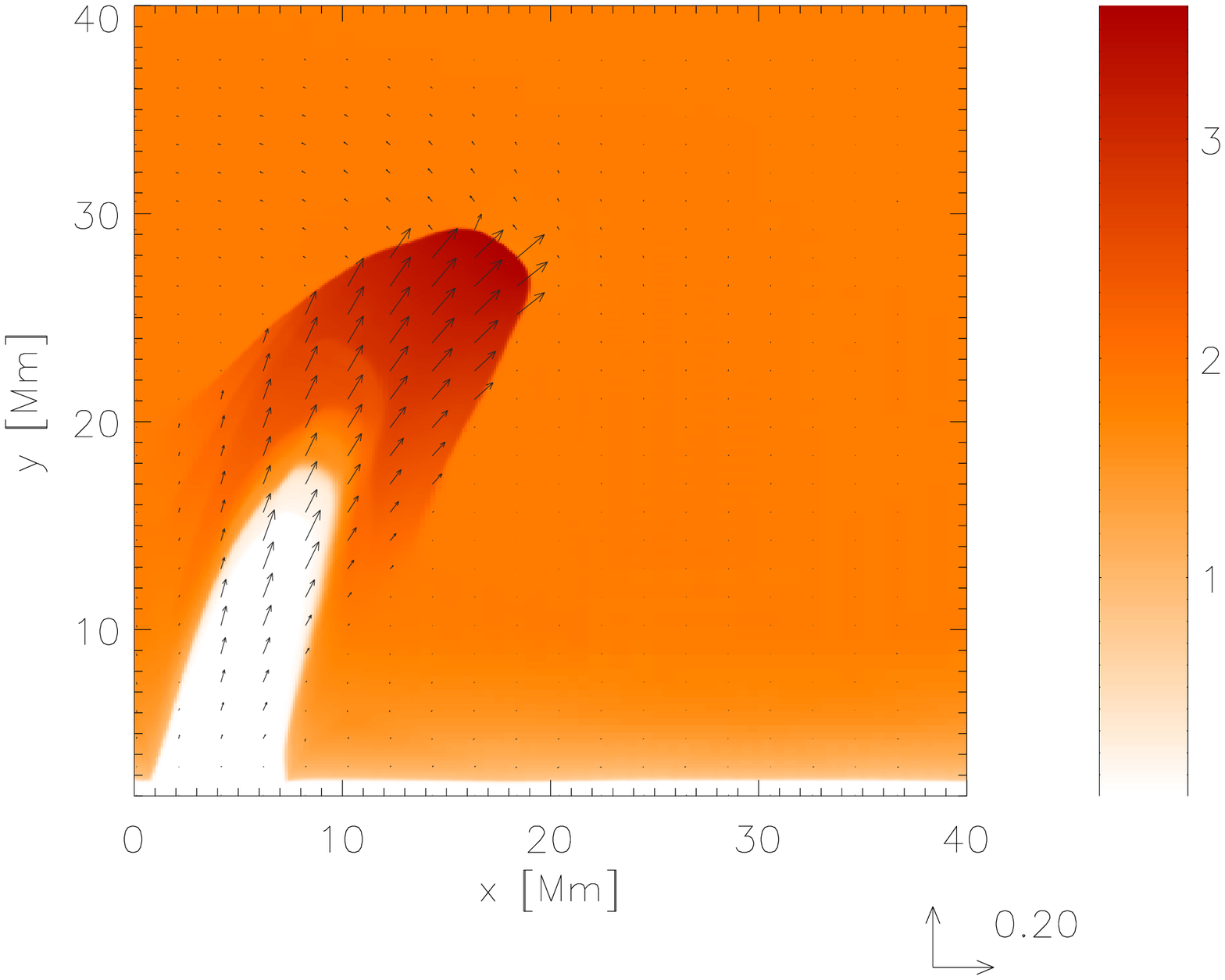}}
\mbox{
\includegraphics[width=5.5cm,height=8.5cm, angle=90]{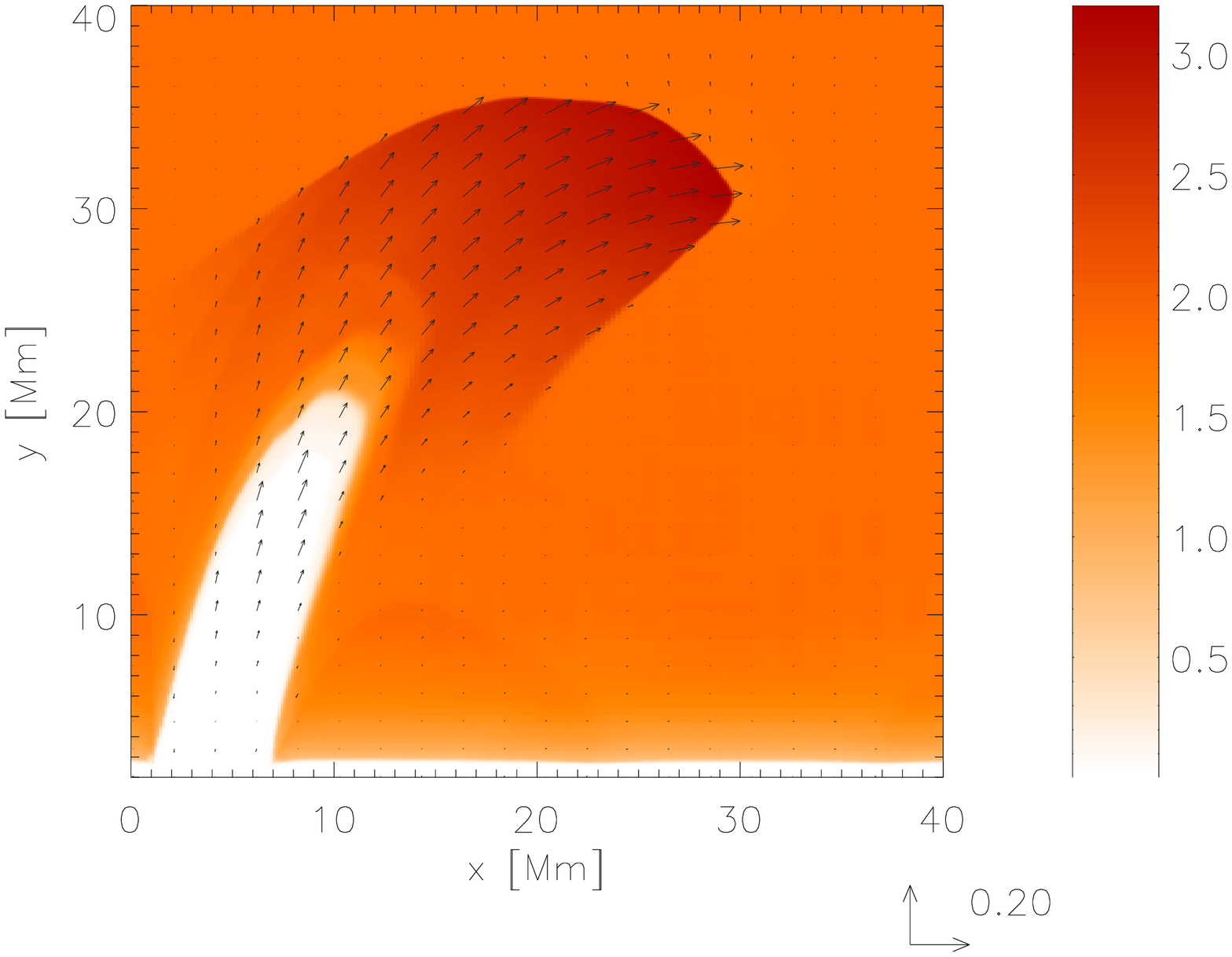}
\includegraphics[width=5.5cm,height=8.5cm, angle=90]{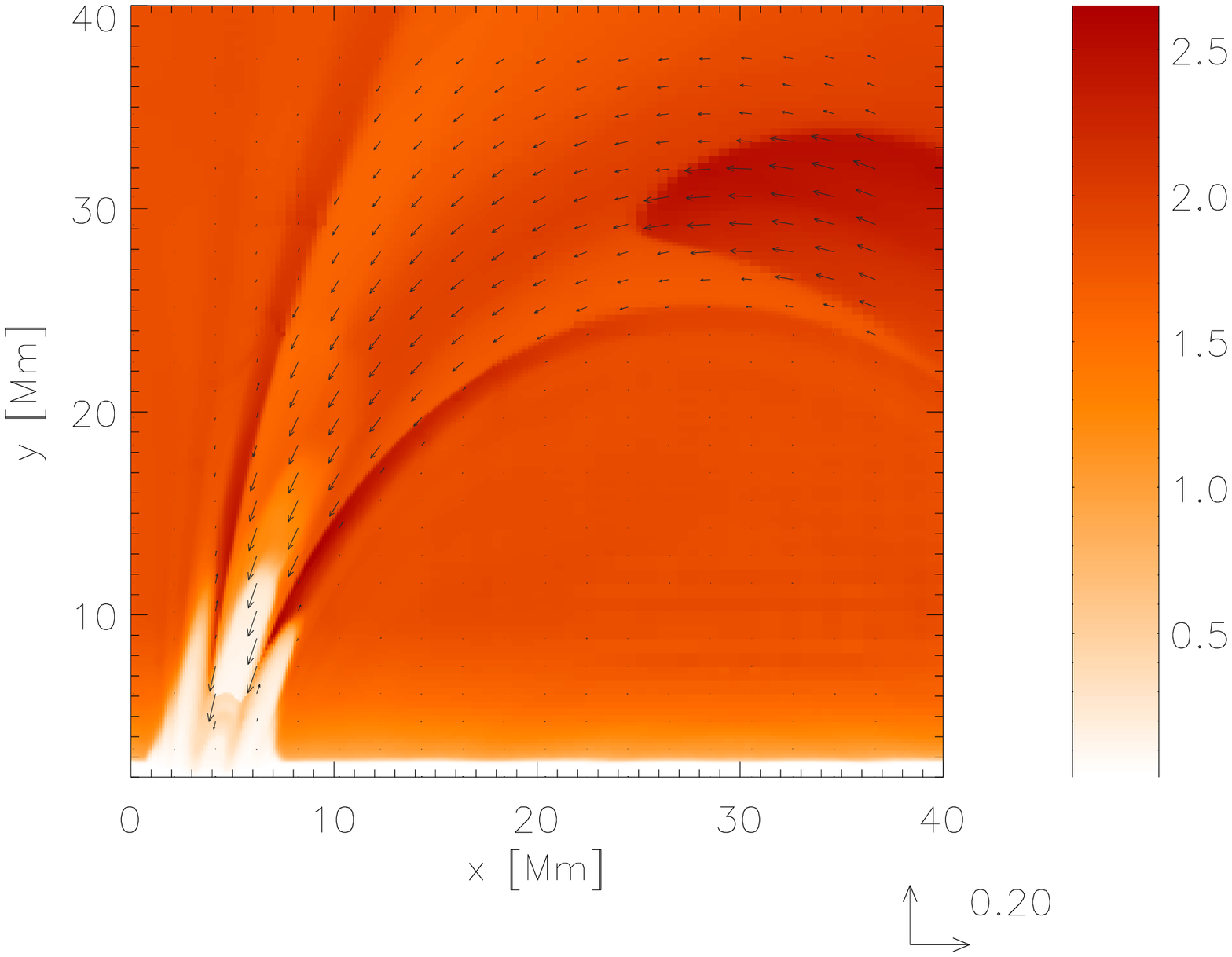}}
\caption{\small 
Numerical results : Temperature (colour maps) and velocity (arrows) profiles at
$t=25$ s, 100 s, 150 s, and 600 s
(from left-top to right-bottom). 
Temperature is drawn in units of $1$ MK. 
The arrow below each panel represents the length of the velocity vector, expressed in units of $200$ km s$^{-1}$. 
}
\label{fig:blob_prof_temp}
\end{figure*}

Equations (\ref{eq:MHD_rho})-(\ref{eq:MHD_B}) are solved numerically using the code FLASH
\cite{Lee2009}. This code implements a second-order unsplit Godunov solver 
with various slope 
limiters and Riemann solvers, as well as adaptive mesh refinement (AMR).
We set the simulation box 
of $(-2.5,65)\, {\rm Mm} \times (2,45.5)\, {\rm Mm}$ along the $x$- and $y$-directions 
and impose fixed in time all plasma quantities at all four boundaries of a simulation region. 
In all our studies we use AMR grid with a minimum (maximum) level of 
refinement set to $4$ ($7$). The refinement strategy is based on 
controlling numerical errors in temperature.
Every block consists of $8\times 8$ identical numerical cells. 

Figure~\ref{fig:blob_prof_temp} displays the spatial profiles of plasma temperature (colour maps) and velocity (arrows) 
resulting from the initial velocity pulse which splits into counter-propagating parts. 
As the plasma is initially pushed upwards the under-pressure results in the region below the initial pulse. This under-pressure sucks up 
cold photospheric plasma which lags behind the shock front. 
The pressure gradient force works against gravity and forces 
the chromospheric material to penetrate the solar corona. 
At $t=25$ s this shock reaches the altitude of $y\simeq 10$ Mm. 
%
The next snapshot (top right panel) is drawn for $t=100$ s. At this time the shock reached the altitude of $y\simeq 30$ Mm 
while the cold plasma blob is located at $y\simeq 18$ Mm. By the next moment of time (bottom left panel) the shock has already moved up 
to the right boundary and the cool counterpart of blob exhibits its developed phase. This is also evident in the H$_{\alpha}$ and TRACE observations that a cool core of 10,000 K and hot coronal plasma maintained at 1-2 MK temperature are simultaneously present in the plasma blob. 
The shock-heated supersonic plasma as well as the cool counterpart both disappear  in the corona at $t=600$ s (bottom right panel). This matches well the observations. 
The blob already subsided 
and the plasma began to flow downward,
being attracted by gravity.
In the present simulation, we only assume the reconnection as a cause that effectively generates 
the radial velocity pulse. We do not invoke the reconnection generated 
joule heating in our model as it can generate the thermal pulse in the ambient medium of certain spatio-temporal
scale that can launch the heated plasma in the upward magnetized atmosphere above the reconnection region \cite{Sri12}.
Therefore, our model takes the flexibility to not consider any initial conditions generated by reconnection, and we
initiate with the launch of the radial velocity pulse triggered due to reconnection process.
We do not invoke therefore either the heating or the losses due to thermal conductivity and radiation 
in our model. The inclusion of these factors will be of the subject of our future study.
\begin{figure*}
\centering
\mbox{
\includegraphics[width=5.5cm,height=8.5cm, angle=90]{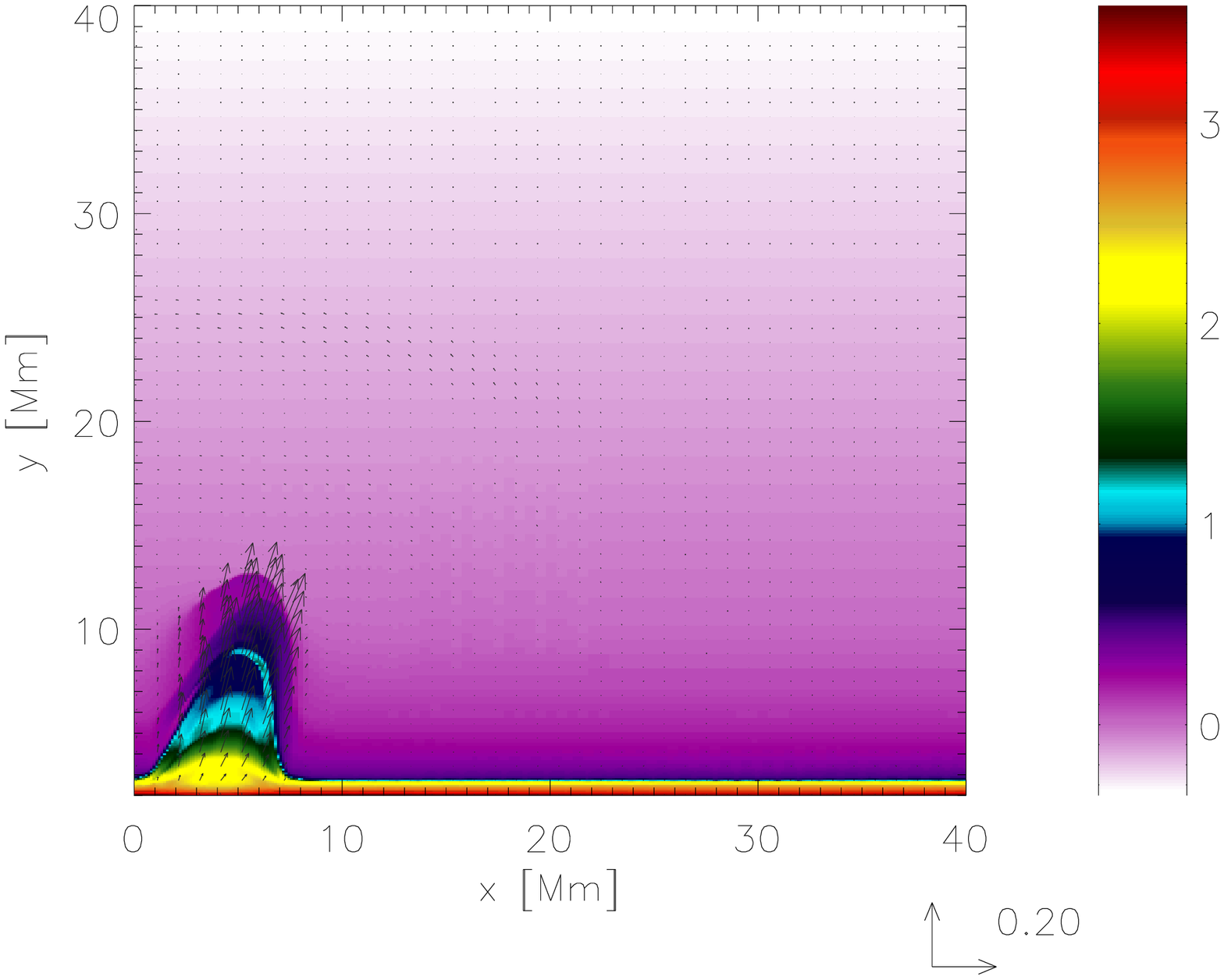}
\includegraphics[width=5.5cm,height=8.5cm, angle=90]{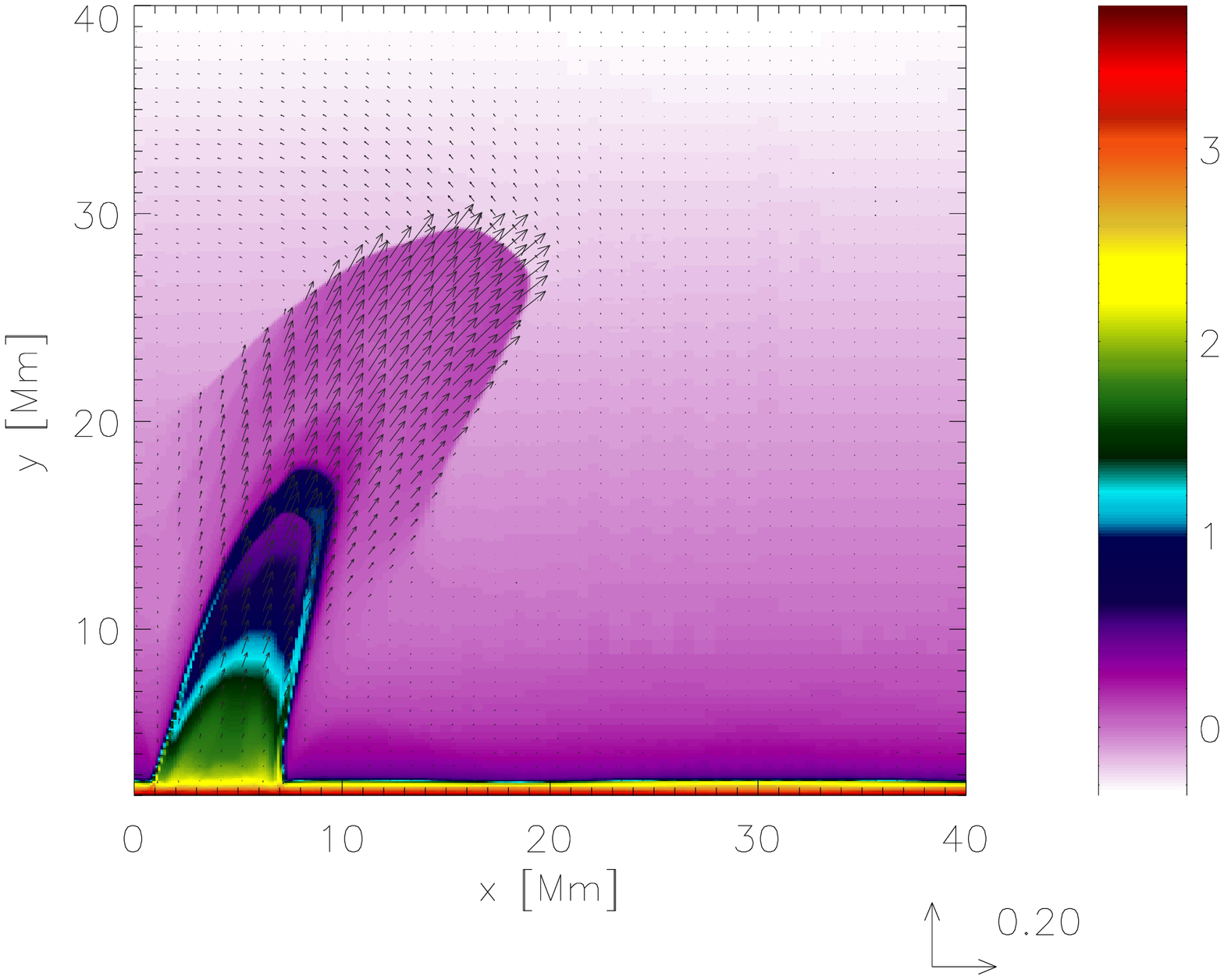}}
\mbox{
\includegraphics[width=5.5cm,height=8.5cm, angle=90]{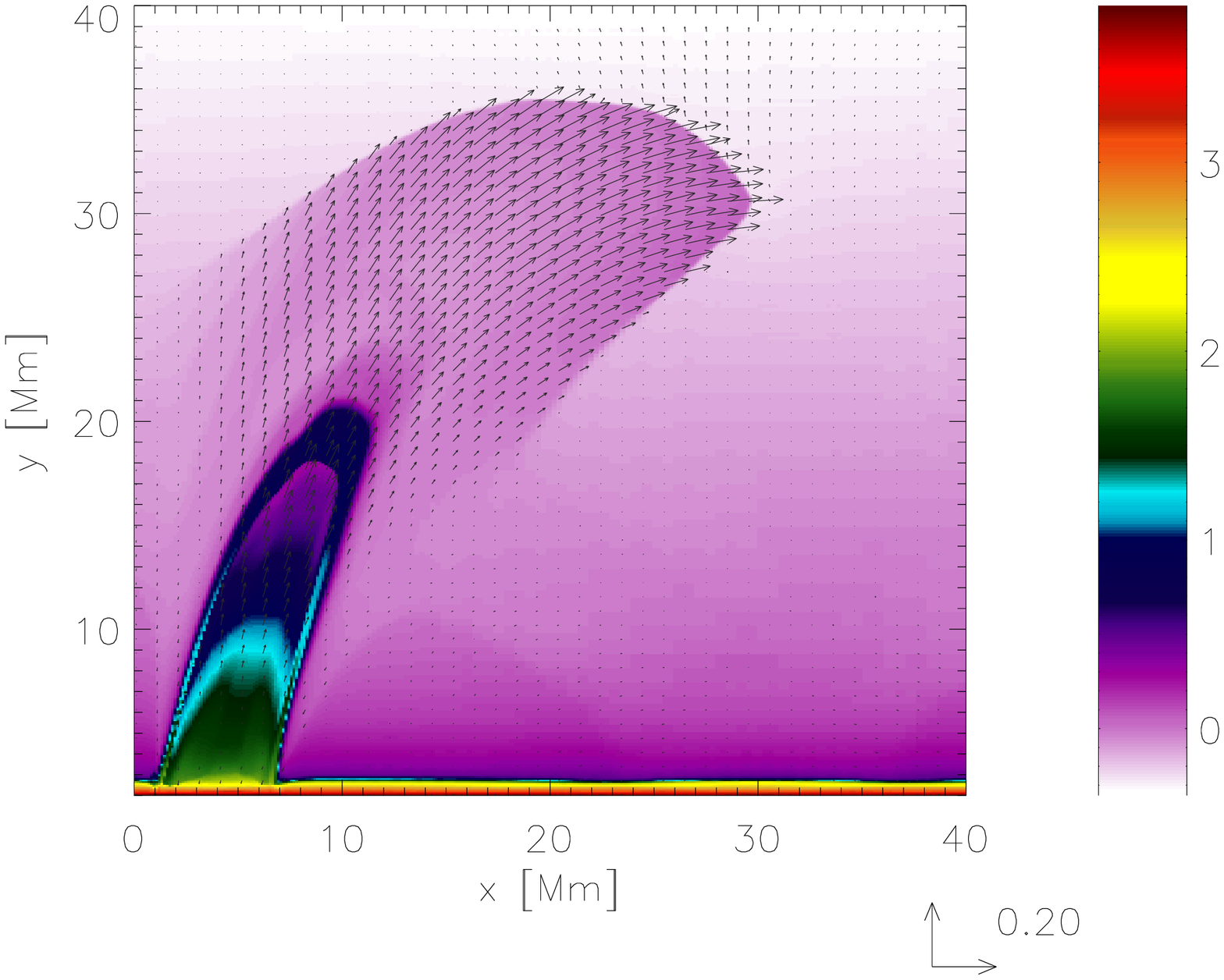}
\includegraphics[width=5.5cm,height=8.5cm, angle=90]{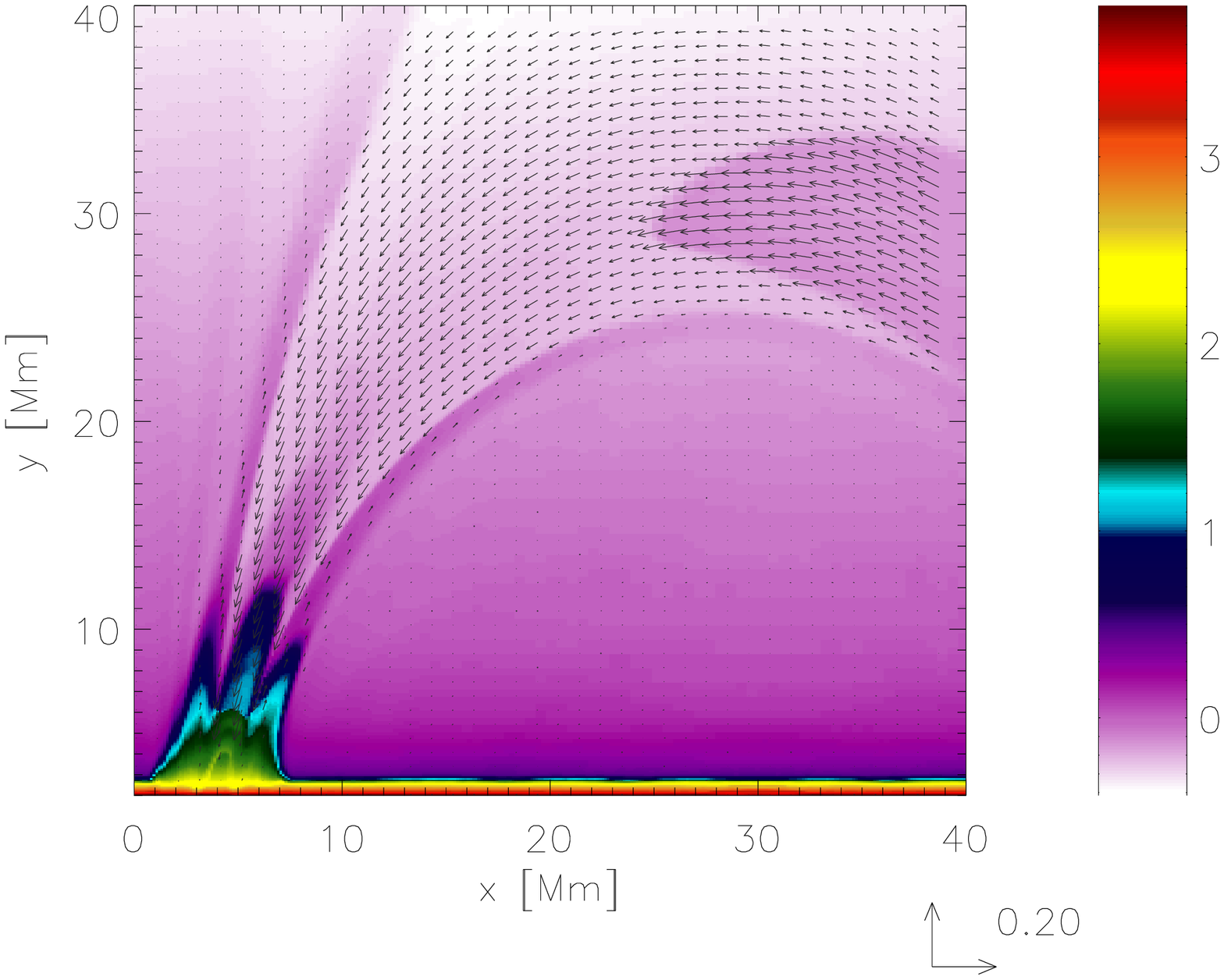}}
\caption{\small 
Numerical results : Density (colour maps) and velocity (arrows) profiles at
$t=25$ s, 100 s, 150 s, and 600 s
(from left-top to right-bottom). 
Density is drawn in logarithemic density scales. 
The arrow below each panel represents the length of the velocity vector, expressed in units of $200$ km s$^{-1}$. 
}
\label{fig:blob_prof_dens}
\end{figure*}

Figure~\ref{fig:blob_prof_dens} displays the spatial profiles of plasma density (colour maps) and velocity (arrows) 
resulting from the initial velocity pulse which splits into counter-propagating parts. 
In the snapshot (top right panel) drawn for $t=100$ s, the shock reached the altitude of $y\simeq 30$ Mm 
while the denser cool plasma blob is located at $y\simeq 18$ Mm. By the next moment of time (bottom left panel) the shock has already moved up 
to the right boundary and the cool counterpart of blob exhibits its developed phase. This is also evident in the H$_{\alpha}$ and TRACE observations that a cool and denser core of 10,000 K and comparatively less denser hot coronal plasma maintained at 1-2 MK temperature are simultaneously present in the plasma blob. These findings of numerical simulation validate our multi-wavelength observations of detached blob shaped jet upto some extent. The shock-heated supersonic plasma as well as cool counterpart both disappear  in the corona at $t=600$ s (bottom right panel) that matches well with the observations. 
The blob already subsided 
and the plasma began to flow downward,
being attracted by gravity.
However, the clear-cut detachment of the blob material
from its origin point is not visible as clearly as in the observations.

\begin{figure}[h]
\begin{center}
\includegraphics[scale=0.45]{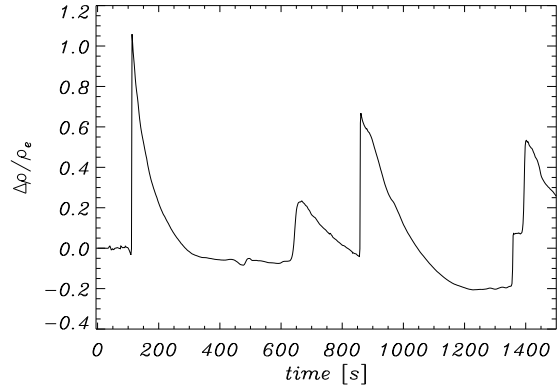}
\caption{\small 
Time-signatures of perturbed mass density 
collected at ($x=20, y=30)$ Mm
for 
the case of Figure 7. 
Time $t$ 
is expressed in units of $1$ s. 
}
\label{fig:time_profile}
\end{center}
\end{figure}

Figure~\ref{fig:time_profile} illustrates the 
relative mass density $(\varrho-\varrho_{\rm e})/\varrho_{\rm e}$. 
that is collected in time at the detection point $(x=20, y=30)$ Mm for 
the case of Figure~7 that mimic the observed blob-shaped detached jet. 
As a result the mass falls off with height and upwardly propagating waves steepen rapidly into shocks. 
The arrival of the first shock front to the detection point
is at $t\simeq 115$ s.
The second shock front reaches the detection point at $t\simeq 865$ s, i.e., after $\sim 750$ s. 
This secondary shock results from the reflected wave from the transition region. 
Therefore, the observation of the intensity (thus density) variations
in the detached phase of the blob as evident in H$_{\alpha}$ and TRACE,
is due to the
variation of density in the plasma blob by the periodic steepening of  
velocity pulse.

%
\section{Discussion and Conclusions}
Using multi-wavelength observations from TRACE 171 \AA\, MDI/SoHO, and H$\alpha$
from Culgoora Solar Observatory, we have observed a supersonic plasma blob 
just before M6.2 class flare in AR 10808 on 9 September 2005. The blob moves 
with a supersonic speed of $\sim$215 km s$^{-1}$ in its dynamical phase and quickly vanishes 
in the corona. The supersonic speed, vanishing nature, increment of the 
intensity (thus density) followed by its decrement during the life-time of blob,
all collectively support the excitation of a velocity pulse 
just in the vicinity of the magnetic null point and reconnection is suggested
to be a primary driver. At the same place, the 
repetitive plasma dynamics at 05:15 UT, and 05:50 UT have also been 
observed. However, they were much fainter and the blob shape plasma
eruption was not evident in those processes. Moreover, those eruptions
were also not visible in the cool H$\alpha$ temperature. Therefore, we have 
a unique observational signature of the propagation of a
supersonic plasma blob during the multiple plasma eruptions
at the flare site in AR 10808. This is the 
most likely clue of the occurrence of the repetitive reconnection 
processes in the observed active region well before and near to
M-class flare, which generates the multiple plasma eruptions as well
as the supersonic blob, which may also be a detached jet that
moves along the spine field lines above the null point.

The velocity pulses launched by the photosphere motions 
and granular power can launch the spicule-like
small-scale plasma jets in the lower solar atmosphere \cite{murawski2010,Mal07}.
However, in the active regions, the reconnection event and 
large energy deposition can generate the strong 
velocity pulses that launch such type of pulse-driven observed plasma blobs.
Therefore, the reconnection-generated velocity pulse is found to be an efficient driver of the 
supersonic plasma blob. 
The energy release by a recurrent 3-D reconnection process via the separator dome below the magnetic 
null point, between the emerging flux and pre-existing field lines in the lower solar atmosphere,
is found to be the driver of a radial velocity pulse outwards that
accelerates this 
plasma blob in the solar atmosphere.
The observed magnetic field and its extrapolation at
the flare site mimic the formation of a 3D null point and the reconnection
through the separator dome that are well established in theory \cite{asc2004}.
Our numerical modelling shows the formation of a supersonic plasma blob with a speed of $\sim$215 km s$^{-1}$
that is triggered due to the generation of a Gaussian velocity pulse at a height of
5 Mm from the photosphere where the low-lying quadrapolar loop system probably reconnects (cf., Figs 1,2,5). The pulse
steepens into shock with the evolution of temperature gradient, and the low
pressure behind it causing the uplift of cool low atmospheric plasma in the 
corona. The observation of multi-temperature plasma in the blob matches well
with this evolution of the multi-temperature plasma in various parts of the 
simulated plasma blob, i.e., the cool confined material enveloped by a heated plasma.
The observed supersonic blob powered by reconnection-generated velocity pulse may be very significant
in fulfilling the coronal energy losses above an active region, if they
occur repeatedly in the corona. This is the unique observational evidence
of such high-speed energy packets which may occur under the particular
magnetic field configuration in the flaring regions. If we assume
the typical coronal electron density above active region loop-like field configuration $n_{e}$=1.0$\times$10$^{10}$ cm$^{-3}$,
the phase speed $V_{ph}$=215 km s$^{-1}$, and the non-thermal speed 
in form of unresolved mass motions in the coronal plasma as $V_{nt}$=40 km s$^{-1}$,
then such MHD  pulse-driven blobs may carry an energy flux of E$_{f}$=7.0$\times$10$^{6}$
ergs cm$^{-2}$ s$^{-1}$ to submit in the corona, if their origin is non-thermal as 
in this case. This energy transport may be very 
sufficient to contribute to re-balance the coronal losses above active regions.

In conclusions, these first observations provide important clues about the plasma dynamics
driven by a velocity pulse, and its role in 
energy transport 
in the solar atmosphere. 
The rare multiwavelength observations of the blob propagation in the solar corona 
also provides the clues of steady reconnection process 
in the low atmosphere well before the flaring event.
Further multi-wavelength
observational studies will be required to shed more light on the transport of
the mass and energy by such reconnection generated and pulse driven plasma
eruptions.

\begin{acks}
We acknowledge the remarks of the referee during review process of our manuscript.
AKS thanks SP2RC, School of Mathematics and Statistics, The University of Sheffield for
the support of collaborative visit, where
the part of present research work has been carried out.
AKS also acknowledges discussions with
Boris Filippov, M. Opher, E. Verwichte, and to Shobhna Srivastava for her
support and encouragement during the work.
We also acknowledge MDI/SoHO and TRACE observations used
in this study. We also thank Culgoora Solar Observatory at Narrabri, Australia
to provide the H-alpha images. 
RE acknowledges M. K\'eray for patient encouragement
and is also grateful to NSF, Hungary (OTKA, Ref. No. K83133) for support
received. 
The software used in this work was in part developed by the DOE-supported ASC/Alliance Center for
Astrophysical Thermonuclear Flashes at the University of Chicago. 
\end{acks}


\bibliographystyle{spr-mp-sola}
\bibliography{references.bib}



\end{article}
\end{document}